\documentclass[sigconf, edbt, table, authorversion, nonacm]{acmart-edbt2021}

\setcopyright{rightsretained}

\acmDOI{}

\acmISBN{978-3-89318-084-4}

\acmConference[EDBT 2021]{24th International Conference on Extending Database Technology (EDBT)}{March 23-26, 2021}{Nicosia, Cyprus} 
\acmYear{2021}

\usepackage{amsmath}
\usepackage{graphicx} %
\usepackage{url}
\usepackage{enumitem} %
\usepackage{bm} %
\usepackage{upgreek} %
\usepackage{bbding} %
\usepackage{pifont} %

\usepackage{listings} %

\usepackage{tabularx} %
\usepackage{booktabs} %
\usepackage{multirow}
\usepackage{hhline} %
\usepackage{diagbox}
\usepackage{pgfplotstable} %

\usepackage{booktabs}  %

\usepackage{tikz} %
\usetikzlibrary{matrix}
\usetikzlibrary{calc}
\usetikzlibrary{math}

\usepackage{adjustbox}
\def\eox{\unskip\kern 10pt{\unitlength1pt\linethickness{.4pt}$\diamondsuit${}}} %

\usepackage{easybmat} %

\usepackage{rotating}		%

\usepackage{subcaption} %
\usepackage[ruled,noend,linesnumbered]{algorithm2e} %

\DontPrintSemicolon     %
\SetNlSty{}{}{}                %
\SetAlgoInsideSkip{smallskip}   %
\SetAlCapSkip{1.5mm}             %
\SetInd{1.5mm}{1.5mm}             %

\SetAlFnt{\small}			%
\SetAlCapFnt{\small}		%
\SetAlCapNameFnt{\small}

\newcommand{\hide}[1]{} 		%

\hypersetup{                                                            %
	pdfpagemode=UseNone,               %
    pdfstartview=Fit,
    pdfpagelayout=SinglePage,       %
    bookmarks=false,
    bookmarksnumbered=true,
    bookmarksopen=false,             %
    pdftex=true,
    unicode,
    colorlinks=true,       %
    linkcolor=blue,          %
    citecolor=cyan,  %
    filecolor=magenta,      %
     urlcolor=blue           %
}

\crefname{claim}{Claim}{Claims}
\crefname{hypothesis}{Hypothesis}{Hypotheses}

\newtheorem{theorem}{Theorem}[section]	%

\newtheorem{questionW}{Question}
\newtheorem{resultW}{Result}

\newtheorem{hypothesis}[theorem]{Hypothesis}

\newtheorem{example}[theorem]{Example}	%

	\newtheorem{definition}{Definition}		%

{{\textit{Question \arabic{questionW}#1.}
  \stepcounter{questionW}}}
{}
\usepackage{tcolorbox}		%
\tcbuselibrary{breakable,skins}		%

\tcbset{examplestyle/.style={
		enhanced jigsaw,	%
		colback=gray!10,	%
		colframe=gray!10,	%
		arc=0mm,
		boxrule=0pt,		%
		left=1mm,
		right=1mm,
		left skip=-1mm,  %
		right skip=-1mm, %
		top=0pt,		%
		bottom=0pt,		%
		breakable,		%
		parbox = false,		%
		before={\par\pagebreak[0]\vspace{1mm}\parindent=0pt},		
		after={\par\pagebreak[0]\vspace{1mm}\parindent=0pt},				
		bottomrule = 0mm,
		boxsep = 0mm,					%
		topsep at break=0pt,			%
		bottomsep at break=0pt,			%
		pad at break=0mm,
		pad before break=0mm,		
		pad after break=1mm,		
		bottomrule at break=0mm,
		toprule at break=0mm,		
		}}

\tcolorboxenvironment{example}{examplestyle}
\usepackage{footnote}%

\tcbset{qrboxstyle/.style={
		enhanced jigsaw,	%
		colback=gray!20,
		colframe=gray!40,
		arc=0mm,
		boxrule=1pt,		%
		left=1pt,
		right=1pt,
		topsep at break=1mm,			%
		top=1pt,		%
		bottom=0mm,		%
		breakable,		%
		parbox = false		%
		}}

\newcounter{resultboxenv}

\newsavebox{\coloredbgbox}

\usepackage{adjustbox}
\def\eox{\unskip\kern 10pt{\unitlength1pt\linethickness{.4pt}$\diamondsuit${}}}

\makeatletter
\DeclareRobustCommand*\uell{\mathpalette\@uell\relax}
\newcommand*\@uell[2]{
  \setbox0=\hbox{$#1\ell$}
  \setbox1=\hbox{\rotatebox{10}{$#1\ell$}}
  \dimen0=\wd0 \advance\dimen0 by -\wd1 \divide\dimen0 by 2
  \mathord{\lower 0.1ex \hbox{\kern\dimen0\unhbox1\kern\dimen0}}
}
\makeatother

\usepackage[normalem]{ulem} %
\newcommand\hl{\bgroup\markoverwith
  {\textcolor{yellow}{\rule[-.5ex]{2pt}{2.5ex}}}\ULon}

\usepackage{marginnote}

\newcommand{\smallsection}[1]{\vspace{5mm}\noindent\textbf{#1.}} %
\newcommand{\introparagraph}[1]{\textbf{#1.}} %

\renewcommand{\epsilon}{\varepsilon} %
\newcommand{\bigO}{{\mathcal{O}}} %

\usepackage{dsfont}
\usetikzlibrary{bayesnet}

\usepackage{ tipa }

\renewcommand\footnotetextcopyrightpermission[1]{}

\begin{document}

\title{DomainNet: Homograph Detection for Data Lake Disambiguation}

\author{Aristotelis Leventidis \quad Laura Di Rocco \quad Wolfgang Gatterbauer \\  Ren\'ee J. Miller \quad Mirek Riedewald}

 \affiliation{%
  \institution{Northeastern University}
  \city{Boston, MA}
  \state{USA}
}
\email{{leventidis.a, la.dirocco, w.gatterbauer, miller, m.riedewald}@northeastern.edu}

\begin{abstract}
Modern data lakes are deeply heterogeneous in the vocabulary that is used to describe data.
We study a problem of disambiguation in data lakes: 
{\em how can we determine if a data value occurring more than once in the lake has different meanings and is therefore a homograph?}
While word and entity disambiguation have been well studied in computational linguistics, data management and data science, 
we show that data lakes provide a new opportunity for disambiguation of data values since they represent a massive network of interconnected values.
We investigate to what extent 
this network can be used to disambiguate values.

\texttt{DomainNet} uses network-centrality measures on a bipartite graph whose nodes
represent values and attributes to determine, without supervision, if a value is a homograph.
A thorough experimental evaluation demonstrates that state-of-the-art techniques in domain discovery cannot be re-purposed to compete with our method.
Specifically, using a domain discovery 
method to identify homographs has a precision and a recall of 38\% versus 69\% with our method on a synthetic benchmark.
By applying a network-centrality measure to our graph representation, \texttt{DomainNet} 
achieves a good separation between homographs and data values with a unique meaning.
On a real data lake
our top-200 precision is 89\%. 
 
\end{abstract}

\maketitle

\section{Introduction}
\label{sec:intro}

Data lakes are large repositories where the metadata, including table names, attribute names, and attribute descriptions may be incomplete, ambiguous, or missing~\cite{nargesian2019tutorial}. 
Modern data lakes are heterogeneous in many different ways:
semantics, metadata, and data values.
We consider 
the problem
of determining if
a data value (i.e., the value of an attribute in a table) that
appears more than once in the data lake has a single meaning.
A data value with more than one meaning is  a \emph{homograph}.  
We illustrate the data lake disambiguation problem through an example.

\begin{example}
\label{ex:intro}
Consider the small sample of a data lake in \Cref{tab:jaguar_ex}, showing
four tables about different topics.
T1 is about corporate sponsorship for efforts to save at-risk species,
T2 is about populations in zoos, 
T3 is about car imports, and
T4  is about corporate sales.
Without disambiguation, a simple keyword search for Jaguar will return a very heterogeneous set of tuples.

One approach to tackle this problem would be to apply document disambiguation by treating tables as documents.
Such techniques are excellent at discerning topics in natural language documents and using this information to further disambiguate the words.
However, because of the nature of tables that are often used to express relationships between different types of entities and values, distinguishing between a donor table $T1$ and a zoo table $T2$ that contain within them synonyms for animals while also being about very different topics (donations and zoos) is a difficult task.
Distinguishing between car manufacturers $T3$ and corporations $T4$ can be even harder because of the prevalence of numerical values.

Entity resolution and disambiguation methods commonly assume a small set of tables about a small number of entity types (which may have the same or different schemas).
In contrast, in a data lake the values to be disambiguated may appear in hundreds of tables about very different entity types and relationships between them.
The ambiguous values need not be named entities, but may be descriptors or any data value in a table.  
This makes entity resolution inapplicable, but opens up new opportunities to use the large network of values and co-occurrences of values in the lake in new ways.  

\end{example}

\begin{figure}[ht]
    \centering
    \small
    \renewcommand{\tabcolsep}{2pt}
    \begin{tabular}{r|llr|}
          \hline
          \rowcolor[HTML]{EFEFEF} 
     $T1$ & Donor & At Risk  & Donation  \\ \hline
          \rowcolor[HTML]{FFFFFF} 
          & Google & Panda & 1M    \\
          \rowcolor[HTML]{FFFFFF} 
          & Volkswagen  & Puma & 2M  \\
          \rowcolor[HTML]{FFFFFF} 
          & BMW  & Jaguar & 0.9M    \\
          \rowcolor[HTML]{FFFFFF} 
          & Amazon  & Pelican & 1.5M  \\ 
\cline{2-4}

        \end{tabular}%
        \quad\quad
    \begin{tabular}{r|llr|}
          \hline
          \rowcolor[HTML]{EFEFEF} 
$T2$ &         name & locale   & num  \\ \hline
          \rowcolor[HTML]{FFFFFF} 
&          Panda & Memphis & 2    \\
          \rowcolor[HTML]{FFFFFF} 
&          Panda  & Atlanta & 2  \\
          \rowcolor[HTML]{FFFFFF} 
&          Lemur  & National & 20    \\
          \rowcolor[HTML]{FFFFFF} 
&          Jaguar  & San Diego & 8  \\ 
\cline{2-4}
        \end{tabular}%
        \vspace{4pt}
        \newline
    \begin{tabular}{r|lll|}
          \hline
          \rowcolor[HTML]{EFEFEF} 
$T3$&          C1 & C2   & C3  \\ \hline
          \rowcolor[HTML]{FFFFFF} 
&          XE     & Jaguar & UK    \\
          \rowcolor[HTML]{FFFFFF} 
&          Prius  & Toyota &  Japan \\
          \rowcolor[HTML]{FFFFFF} 
&          500  & Fiat & Italy   \\
\cline{2-4}
      \end{tabular}%
      \quad\quad
      \begin{tabular}{r|lrr|}
      \hline
      \rowcolor[HTML]{EFEFEF} 
$T4$ &      Name     & Revenue    & Total      \\ \hline
      \rowcolor[HTML]{FFFFFF} 
&      Jaguar    & 25.80 & 43224 \\
      \rowcolor[HTML]{FFFFFF} 
&      Puma & 4.64 & 13000 \\
      \rowcolor[HTML]{FFFFFF} 
&      Apple  & 456 & 370870 \\
&      Toyota  & 123 & 123456 \\
\cline{2-4}
      \end{tabular}%
\caption{Running example with Jaguar and Puma having multiple
meanings. How can we use co-occurrence information across a data
lake to discern different meanings?} 
\label{tab:jaguar_ex}
\end{figure}

In entity resolution (ER)~\cite{christophides2019end},
the idea is to determine if two (or a set of) tuples 
refer to the same real-world entity or not.
An important assumption in ER is that the tables being resolved are about the same (known) entity types.
As an example, given a set of tables about papers that include authors as data values, we can determine if two tuples refer to the same paper (have the same meaning).
As a by-product of entity resolution, a data value, for example ``X. Wang,'' may be identified as an ambiguous data value that refers to more than one real-world entity. 
Schema-agnostic ER techniques have  been proposed that do not assume the entities are represented by the same schema~\cite{papadakis2020three}.
However, these approaches still assume the tables being resolved represent entities of the same type.

In our problem, we are not starting with a small set of tables that are known to refer to the same type of real-world entities, e.g., customers or research papers.  
We want to understand in a data lake with a massive number of tables if the value ``Puma'' in T1 (see \Cref{tab:jaguar_ex}), Attribute \texttt{At Risk} refers to the same real-world concept (not necessarily an entity) as ``Puma'' in Table T4, Attribute \texttt{Name}.

Disambiguation of words in documents has also been heavily studied~\cite{iacobacci2016embeddings,BGB04,SG14,yadav2018survey}.
Solutions often rely on language structures or labeled training data.
In contrast to documents, which are free text, tables are structured and lack the same intuitive notion of \emph{context}.
While plenty of research has explored disambiguation of documents, to the best of our knowledge there is no work on disambiguation of data lakes.
This is of  importance because data lakes can contain many data values that have different meanings. 
As an example, ``Not Available'' is a well known way to represent NULL values in a table.
``Not Available'' is not ambiguous from a natural-language point of view.
However in a data lake it may appear in multiple attributes corresponding to names, telephone numbers, IDs etc., making ``Not Available'' a homograph meaning ``unknown name'' or ``unknown number,'' etc.

Determining if a value in a data lake has a single or multiple meanings
is unexplored territory. We define data lake disambiguation as follows:
\begin{definition}[Data lake Disambiguation]
\label{def:homograph}
Given a data lake containing a collection of tables with possibly missing, incomplete, or heterogeneous table and attribute names.
For any data value $v$ that appears in more than one attribute (column) or table, determine if it has a single meaning or more than one meaning.
The latter are called homographs.
\end{definition}

A homograph is not necessarily a single word from a dictionary or a vocabulary.
In a data lake, a homograph can be a phrase, initialism (e.g., ``NA''), identifier, or any blob (data value).
We do not assume homographs to be named entities; they can be adjectives or another part of speech.  
Homographs arise naturally from words used in different contexts, e.g.,
the classic example of \emph{Apple} as a fruit or a company, 
or \emph{Jaguar} in \Cref{ex:intro}.  
They can also arise due to errors, e.g., when animal color ``yellow'' is accidentally entered in the habitat column. We consider this now ambiguous value
a homograph. Notice that updates to the data lake can change a homograph
to a value with a single meaning, e.g., when the table with the only
alternative meaning is removed; and vice versa.

In this work, we examine the global co-occurrence of data values within a data lake 
and how such information can be used to disambiguate data values.  
We show that a local measure is not sufficient and motivate why and how the full network of value co-occurrences enables effective disambiguation.  
This network exploits table structure and had not been considered in the most commonly studied disambiguation problems such as named-entity disambiguation and entity resolution. 
Its disambiguation power comes at a price: The value co-occurrence information
is massive and it is not obvious how to process it efficiently for disambiguation.

\introparagraph{Contributions} 
We address the data lake disambiguation problem using a network-based approach called \texttt{DomainNet}.
Our main contributions are as follows.
\begin{itemize}[leftmargin=*, itemindent=0pt]
  \item We define the problem of homograph detection in data lakes.
  Homographs may arise in tables that do not represent the same (or even similar) types of entities, and hence cannot be identified using entity resolution and disambiguation. 
  They may not even be words in natural language and do not appear in
  natural-language contexts, making language models ineffective.
  \item We present \texttt{DomainNet}, a network-based approach to determine if a data value appearing in multiple attributes or tables is a homograph.
  \texttt{DomainNet} is motivated by work on community detection where a community represents a meaning for a value (e.g., animal or car model).  A homograph is then a value that occurs in multiple communities.
   However, in the homograph detection problem ($i$) there are an \emph{unknown} and possibly \emph{large} number of meanings for a value 
   and
   ($ii$) our goal is to find \emph{values} that span communities, not the communities.
  We identify 
  two measures for finding such community-spanning values, the {\em local clustering coefficient}~\cite{watts1998collective}  
  and the {\em betweenness centrality}~\cite{10.2307/3033543}, and empirically evaluate  their usefulness in homograph detection.  
 \item We present an evaluation on a synthetic dataset (with ground truth),
 studying the performance of both centrality measures and motivating the use of the more computationally expensive betweenness centrality. 
 We compare \texttt{DomainNet} to a recent unsupervised domain detection algorithm $D^4$~\cite{Ota+20} (any value belonging to multiple domains is a homograph).
 $D^4$ achieves a precision and a recall of 38\% whereas \texttt{DomainNet} reaches 69\%.
   \item We create 
a disambiguation benchmark from the real data used in a recent table-union benchmark~\cite{nargesian2018table} and show that we can effectively find
   naturally occurring homographs in this data (89\% of the first 200 retrieved values are homographs based on ground truth). 
   We also systematically introduce homographs into real data and show that betweenness centrality achieves 85\% accuracy when homographs are injected into both small and large attributes, and over 97\% accuracy when homographs are all injected into attributes with at least 500 distinct values.
   We show that \texttt{DomainNet} is effective even when there is high variance in the number of meanings of different homographs.
   \item
   To illustrate the importance of homograph discovery, we show the impact that as few as 50 homographs (injected into a clean unambiguous real data lake) can have on a domain discovery algorithm~\cite{Ota+20}.
   As the number of of homographs increases, the accuracy of the domain discovery algorithm deteriorates. 
  \item 
 The scalability of our approach depends on the size of the data lake vocabulary (the number of values) and on the density of the network (number of edges).
  We use real data (from NYC open data) with a vocabulary size of 1.5M to show that 
  we can compute the \texttt{DomainNet} network in 3.5 min and find homographs in 27 min
  using an approximation of betweenness centrality based on sampling. 
\end{itemize}

The remainder of this paper is organized as follows.
In \Cref{sec:related} we discuss existing work in disambiguation.
In \Cref{sec:approach+networks} we introduce our approach and describe how applying centrality measures on a graph representation of the data lake can be used to identify homographs.
\Cref{sec:setup} summarizes the datasets used in our experimental evaluation presented in \cref{sec:experiments}.
We conclude and outline possible future directions of our work in \Cref{sec:conclusion}.
For further information, please visit our project page at \url{https://northeastern-datalab.github.io/table-as-query/}

\section{Foundations of Disambiguation}
\label{sec:related}

Disambiguation has been studied in several contexts in NLP, data management and broadly in AI and data science. We analyze how this work can be applied
to disambiguation in data lakes.

\subsection{Entity Resolution}
Entity Resolution (ER) identifies records (also called tuples) across different datasets (or sometimes corpora) that represent the same real-world entities.  
ER is generally applied to structured and semi-structured data including tables and RDF triples~\cite{GM12}.
Some ER approaches also identify ambiguous values as part of the resolution process.
For example, using collective entity resolution over two types of tables (e.g., papers and authors) one can identify if a value, say ``X. Wang,'' refers to different authors~\cite{BG07}.
Similarly in familial networks, one can resolve synonyms (different values that refer to the same person) and identify homographs (same value used to refer to different people)~\cite{Kou+19}.

ER assumes that the information to be resolved or disambiguated is of a single known type (e.g., resolving customer tuples or patient records) or a small set of types (e.g., authors, their papers, and publishing venues).
Some work, called schema-agnostic ER, does not require that all data be represented using the same schema~\cite{christophides2019end}.
However, all these approaches start with the assumption that two or more tables (or corpora) are describing the same type of entities~\cite{papadakis2016comparative, simonini2016blast,papadakis2020three}.

In data lake disambiguation, we seek to find ambiguous values even when we do not know what type of entities a table is describing.
We also do not know if different tables are describing the same or different entities.
Hence, we cannot apply collective models or other resolution models that rely on this knowledge.

\begin{example}
Given the four tuples with \texttt{Jaguar}: \texttt{[BMW, Jaguar, 0.9M]}, \texttt{[Jaguar, San Diego, 8]}, \texttt{[XE, Jaguar, UK]}, and \texttt{[Jaguar, 25.8, 43224]},
does \texttt{Jaguar} have the same meaning?
These four tuples correspond to four different types of facts: donors and the amount they contribute to protect an endangered species, animals in zoos, car models, and economic information about companies.
ER schema-agnostic algorithms are insufficient in resolving (or disambiguating) values within these heterogeneous tables because they rely on the hypothesis that the tables they examine refer to the same type of real-world entity.
\end{example}

\subsection{Semantic Type Detection}

A possible approach to data lake disambiguation is to discover semantic types for all
attributes (columns) and then label a value appearing in different semantic types
a homograph.
In the running example, identifying the semantic type of \texttt{T1.At Risk} and \texttt{T2.name} as animal and mammal, respectively, and knowing that mammals
are animals, one can infer that Jaguar is not a homograph there.
In contrast, recognizing \texttt{T3.C2} is of type ``Car Manufacturer,'' which is
neither a sub- nor super-type of animals, implies that Jaguar in \texttt{T3} and
\texttt{T1} represents a homograph.
Here, we discuss different approaches to semantic type discovery and to what extent they could be used for homograph detection.  

\introparagraph{Knowledge-based Techniques}
There has been considerable work on semantic type detection in the Semantic Web community that uses external knowledge from well-known ontologies including DBpedia~\cite{DBpedia:2014}, Yago~\cite{Suchanek:2007} and Freebase~\cite{bollacker2008freebase}.
Most solutions have been applied to Web tables~\cite{Lehmberg:2016,Eberius+13,Eberius:2015}
that are small (in comparison to other data lakes) and have rich metadata (table and attribute names).

Hassanzadeh et al. \cite{HWRS15} use a map-reduce approach to find similarity between a (column, data value) pair from a table with a (class, instance label) pair from the Knowledge Base (KB).
Ritze et al.~\cite{RLOB16} match Web tables to DBpedia to profile the potential of Web tables for augmenting knowledge bases with missing information.
These approaches cannot infer type information for an attribute that it is not part of
the KB. Unfortunately, the coverage of values from data lakes in Open KBs is low
(a recent study reports about 13\%~\cite{nargesian2018table}), limiting their applicability.

\introparagraph{Supervised Techniques}
An alternative are machine learning (ML) techniques that infer the semantic type of
attributes. ML solutions utilize a variety of graphical models 
(Conditional Random Fields~\cite{goel2012exploiting}, Markov Random Fields~\cite{Limaye:2010}), as well as Multi-level Classification~\cite{takeoka2019meimei}, and Deep Learning~\cite{Hul+19}.
Sherlock~\cite{Hul+19} uses features about the values in an attribute to classify some of the attributes in a data lake into one of 78 semantic types (like address or horse jockey)~\cite{Hul+19}.
A recent solution, called SATO~\cite{Zha+19}, augments this approach and shows that using
row information can improve the classification accuracy for the same 78 semantic types. 
These approaches require large amounts of labeled training data and are limited by
the set of pre-defined types.  

\introparagraph{Unsupervised Techniques}
Unsupervised semantic type discovery algorithms have only recently started to be studied.
We discuss two unsupervised algorithms, one for semantic type discovery, $D^4$~\cite{Ota+20}, and one for table unionability search~\cite{nargesian2018table}.

$D^4$ provides an unsupervised approach with a focus on assembling all the values of each semantic type in a data lake~\cite{Ota+20} (these values are called a "domain").
They propose a data-driven approach that leverages
value co-occurrence information to cluster values that are from the same domain.
Heuristics attempt to deal with ambiguous values that may appear in multiple domains.  
In our context, $D^4$ can be used to label values that appear in multiple domains as homographs. This indeed serves as a baseline in our experiments.  

Table Union Search~\cite{nargesian2018table} solves a different problem.
Given a query table, they find a set of tables from the lake that are most unionable with
it. In order to do so, they provide several similarity measures that are used collectively
to calculate 
how unionable two attributes are.  
This work can use both ontological and semantic (word embedding) signals when present to determine unionability  over heterogeneous attributes, but does not attempt to find or label homographs.  

\subsection{Disambiguation in Related Areas}

Word-sense disambiguation (WSD)~\cite{navigli2009,iacobacci2016embeddings}, i.e.,
the task of identifying which meaning of a word is used in a sentence,
is an important problem in computational linguistics.
Although a human can proficiently perform this task on a document, constructing algorithms that perform this task effectively is still an open research problem.
Techniques proposed so far range from dictionary-based methods, which use the knowledge encoded in lexical resources (e.g., WordNet)~\cite{navigli2009}, to more recent solutions in which a classifier is trained for each distinct word on a corpus of manually sense-annotated examples~\cite{PilehvarN14}.
Additionally, completely unsupervised methods have also been proposed that cluster occurrences of words, thereby inducing word senses, i.e,  word embeddings~\cite{iacobacci2016embeddings}.
The aforementioned solutions rely on information (or latent information) about the structure of sentences including grammatical rules.
Finally, while solutions that do not rely on grammar also exist, they only operate on documents and not tables~\cite{BGB04,SG14}.

Another relevant sub-task in Natural Language Processing is Named-Entity Recognition (NER), which has been proposed as a possible solution for disambiguation~\cite{yadav2018survey}.
NER seeks to locate and classify named entities mentioned in unstructured text into pre-defined categories such as person names, organizations, locations, etc.
NER systems have been created that use linguistic grammar-based techniques as well as statistical models~\cite{agerri2016robust}.

A special case of the NER problem is the author name disambiguation problem~\cite{smalheiser2009author,ferreira2012brief}
Authors of scholarly documents often share names which makes it hard to distinguish each author's work.
Hence, author name disambiguation aims to find all publications that belong to a given author and distinguish them from publications of other authors who share the same name.
Different solutions have been proposed using graphs~\cite{levin2010evaluating}.
However, the graph structure proposed is largely domain specific.
The graph contains not only the information about the co-authorship and published papers, but also venue of the paper published, year of research activities and so on.

\section{Disambiguation using \texttt{DomainNet}}
\label{sec:approach+networks}

We now present our proposed solution, \texttt{DomainNet}\footnote{The code for \texttt{DomainNet} and our benchmarks is available at \url{https://github.com/northeastern-datalab/domain_net}.}, for finding homographs in a data lake.

\subsection{Problem Definition}
\label{sec:probdef}

Recall from \Cref{def:homograph} that a {\em homograph} is a data value that appears in at least two attributes with more than one meaning.
Values that are not homographs are {\em unambiguous values}.
In data lakes, attribute and table names can be missing or misleading
(with many ambiguous terms like ``name,'' ``column 2,'' or
``detail'')~\cite{nargesian2019tutorial}. Well-curated enterprise lakes may have more
complete metadata, but even they do not follow the unique name assumption---which states that different attribute names always refer to different things.
As a result, many data lake search approaches rely solely on the table contents~\cite[and others]{Zhu:2016,DBLP:conf/icde/FernandezMNM19,DBLP:journals/corr/abs-2010-13273}. In a similar vein, in \texttt{DomainNet}, we investigate to what extent data values and the co-occurrence of data values within attributes can be used to determine if a value is a homograph.

\begin{figure}[!ht]
    \centering
    \includegraphics[width=\columnwidth]{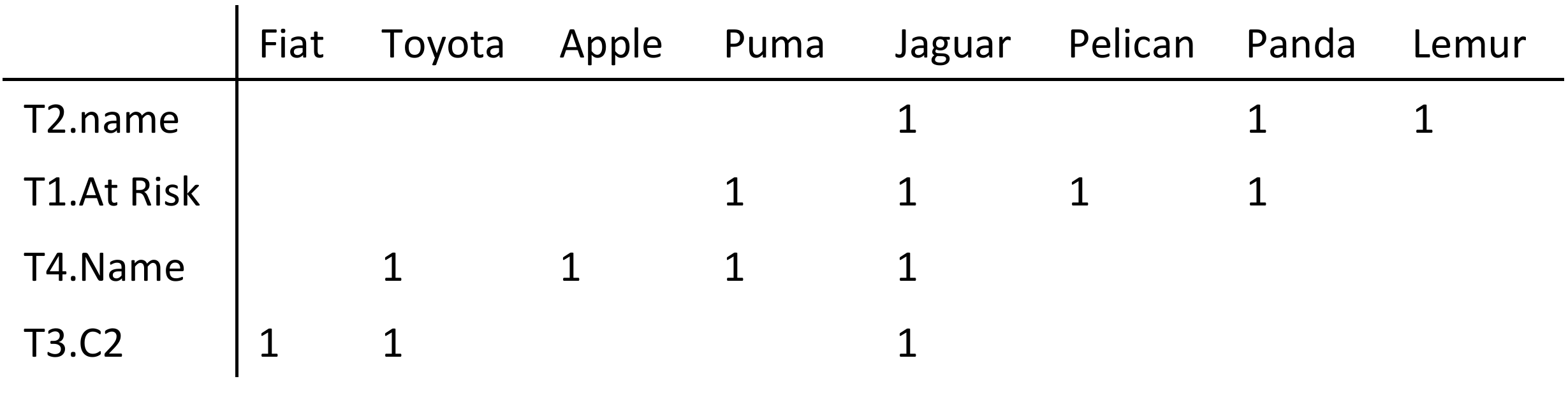}
    \caption{Incidence matrix: vertical axis attributes, horizontal axis data values.
    }
    \label{fig:Example_Fig_a}
\end{figure}

\begin{example}
\label{ex:homograph}
In \Cref{tab:jaguar_ex},
the data value \texttt{Jaguar} is a homograph because it refers to the animal in Tables $T1$ and $T2$ and refers to the car manufacturer in Tables $T3$ and $T4$.
Other values such as \texttt{Panda} and \texttt{Toyota} are unambiguous since they only have a single meaning across \emph{all} tables.
\texttt{Puma} is also a homograph, appearing as an animal and a company.
\Cref{fig:Example_Fig_a} displays which values co-occur with \texttt{Jaguar} in the same column using an incidence matrix: 
the vertical axis shows the different values, and the horizontal axis the different attributes occurring in the data lake.

Note that homographs need not be values from a dictionary. They can be any data value
that appears in a table. Another example of a homograph is the data value \texttt{01223} which in some attributes may refer to a  Massachusetts zip code and in others to an area code near Cambridge, UK, and in yet others to the suffix of an Oil Filter Element Replacement product code. 
\end{example}
In a well-curated database or warehouse, we may know the semantic meaning of each attribute (e.g., %
"Animal Name" vs. "Company Name") and can leverage it to identify homographs.
However, in a dynamic, non-curated 
data lake, we cannot rely on this information to be available.

\subsection{\texttt{DomainNet}: Viewing Values as a Network}\label{subsec:domain_net_network}

In data lakes, without {\em a priori} knowledge of table semantics or types, 
we take a network-based approach to understanding the meaning of repeated data values.  
We propose to detect homographs using network measures. 
For that purpose, we can interpret the co-occurrence information about values across different attributes 
using a network representation in which nodes represent data values and edges represent the fact that two values co-occur in at least one column (attribute) in the data lake.

\begin{figure}[!ht]
    \centering
    \begin{subfigure}[b]{0.45\columnwidth}
        \centering
        \includegraphics[height=3cm]{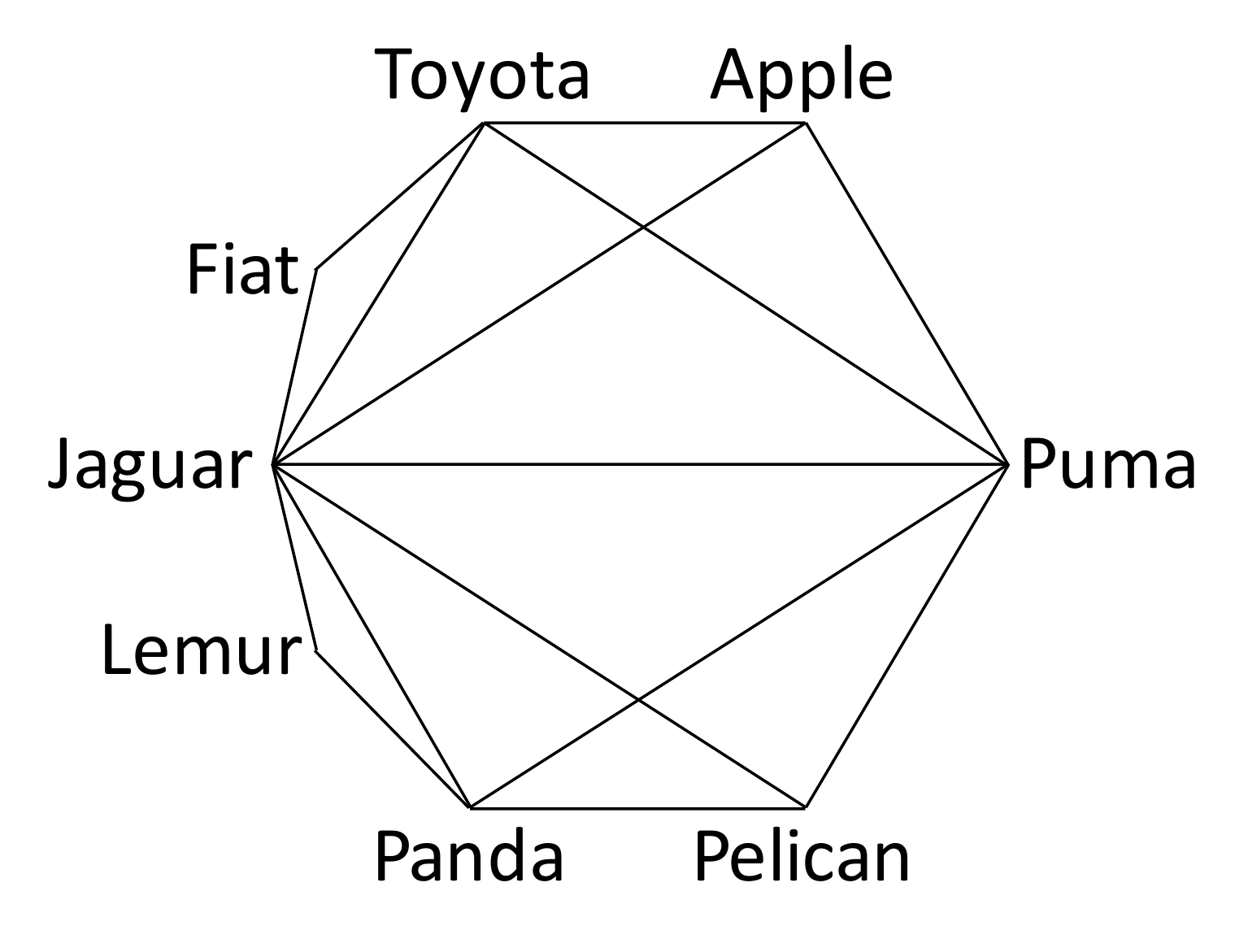}
        \caption{Co-occurrence graph}
        \label{fig:Example_Fig_b}
    \end{subfigure}	
    \begin{subfigure}[b]{0.45\columnwidth}
        \centering
        \includegraphics[height=3cm]{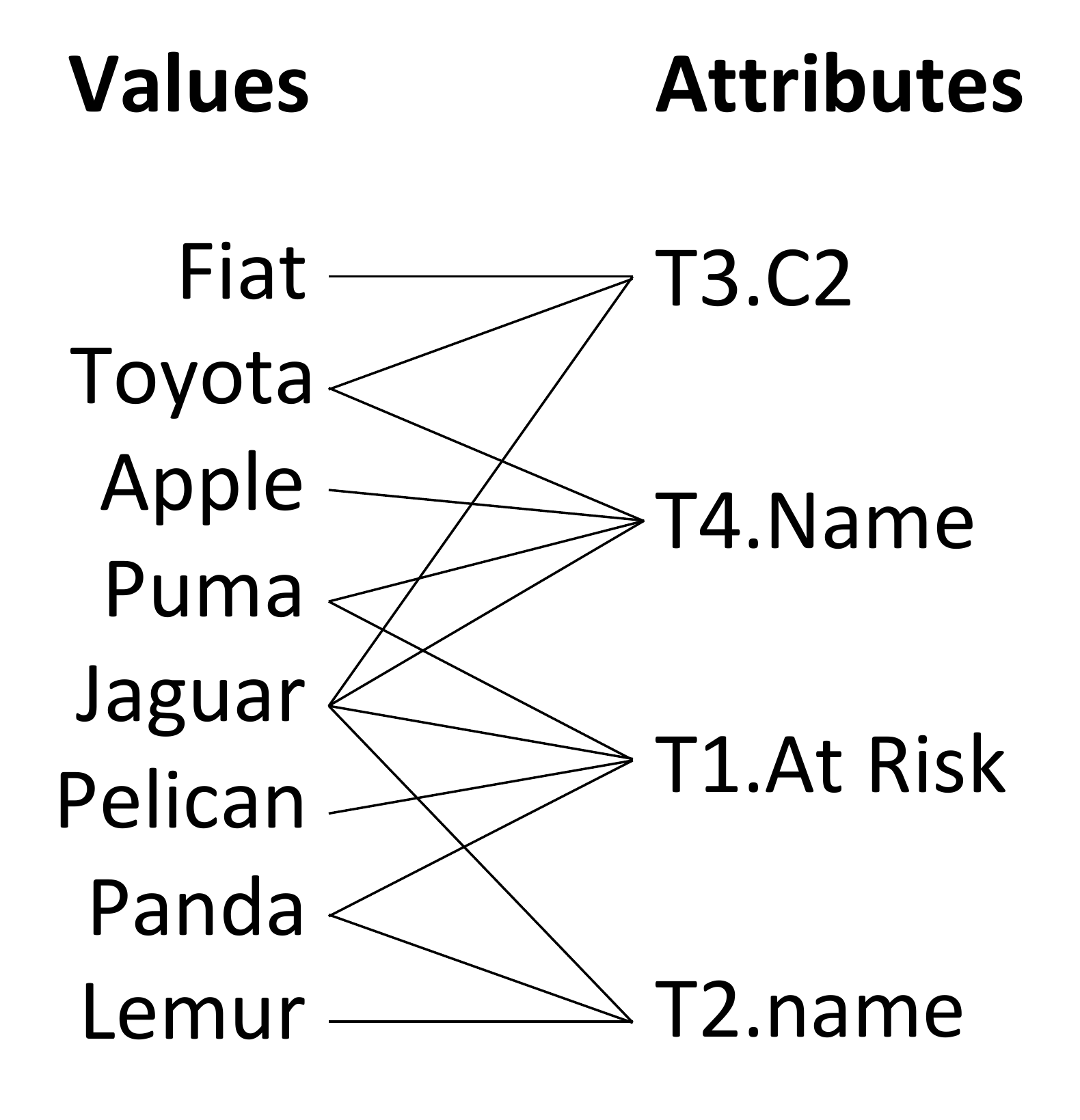}
        \caption{Bipartite graph}
        \label{fig:Example_Fig_c}
    \end{subfigure}	
  \caption{Two graph representations of a portion of \Cref{tab:jaguar_ex}.}
  \label{fig:Example_Fig}
\end{figure}

\begin{example}
    \label{ex:co-occurrence-network}
In \Cref{fig:Example_Fig},
we depict the values from the same four attributes shown in \Cref{ex:homograph}.
\Cref{fig:Example_Fig_b} shows the value co-occurrence network.
Notice that by removing both ``Puma'' and ``Jaguar'' the remaining nodes become \emph{disconnected} into two components.
This captures the intuition that those two values are pivotal in that they bridge two otherwise disconnected meanings or graph components.
\end{example}

Whereas this representation allows us to apply straightforward metrics from community detection, it comes at a high cost: 
the representation uses more space than the original data lake.
The incidence matrix is sparse and has as many entries as there are cells in the data lake (\Cref{fig:Example_Fig_a}).
In contrast, the co-occurrence graph increases quadratically in size with respect to the cardinality of attributes (the size of the vocabulary) in the data lake (\Cref{fig:Example_Fig_b}).
Consider a 
single column with 100 values. The incidence matrix represents this information with 100 rows, 1 column, and 100 entries.
The co-occurrence graph represents this with 100*99/2=4950 edges across 100 nodes.

Thus, we 
use
a more compact network representation that allows us (after some modifications) to apply 
network metrics to discover pivotal points (\Cref{fig:Example_Fig_c}).
\texttt{DomainNet} uses a bipartite graph composed of (data) value nodes and attribute nodes.
The attribute nodes represent the set of attributes and the value nodes the set of data values across all attributes in the lake.
Every data value is treated as a single string, it is capitalized and has its leading and trailing white-space removed to ensure consistent comparison of data values across the lake.
Notice that each data value, even if found in multiple attributes, is represented by one single value node in the graph.
An edge is placed between a value node and an attribute node if the data value appears in the attribute (column) corresponding to that attribute node.  
Data values that appear in more than one attribute are candidates for being homographs. 

\begin{example}
    \label{ex:jaguarNetwork}
\Cref{fig:Example_Fig_c}, shows a portion of the \texttt{DomainNet} representation for \Cref{tab:jaguar_ex} using only the four attributes of ~\Cref{ex:homograph}.
\end{example}

In the \texttt{DomainNet} bipartite graph, we call two data values {\em neighbors} if they both appear in the same attribute (and hence there is a path of length two between them in the graph).
Similarly, two attributes are {\em neighbors} if they have at least one data value in common (and hence there is a path of length two between them).
For a data value node $v$, $N(v)$ denotes the set of all its value neighbors.
We also define the \emph{cardinality of a data value} node $v$ as the number of neighbors $|N(v)|$, which is the number of unique data values that co-occur with $v$.
If $n$ is the number of value nodes and $a$ the number of attribute nodes, the number of edges in a \texttt{DomainNet} graph over real data tends to be much less than $n\cdot a$.

\introparagraph{Tables to Graph}\label{par:graphs}
Recent work on embedding algorithms in relational databases~\cite{KoutrasFKL20,CappuzzoPT20,abs-2005-06437} use a graph representation of tables.
Like \texttt{DomainNet}, they model values and columns as nodes. Depending on the problem addressed, some approaches also include nodes for rows and tables.
Like in our approach, column names are not assumed to be present or unambiguous.  

Koutras et al.~\cite{KoutrasFKL20} and Capuzzo et al.~\cite{CappuzzoPT20} use a tripartite graph representation in which every value node is connected with its column node and its row node.
Such an approach works well for the tasks of tuple-level entity resolution and for schema matching (a similar task to semantic type discovery).
We experimented using both row and table information in \texttt{DomainNet} and found it was not useful in disambiguating values.
In our example, \texttt{Panda} in $T1$ and $T2$ are not homographs, but the row information makes them seem quite different and we did not find it helpful.  

In contrast, Arora and Bedathur~\cite{abs-2005-06437} 
use a homogeneous graph using only data value nodes that are connected with each other if they appear in the same row of the table.
They do not use the value co-occurrence information within a column, making homograph
detection using solely row context inappropriate in large heterogeneous datasets.

\subsection{Homograph-Disambiguation Methodology}\label{subsec:homograph_disambiguation_methodology}

Intuitively, data values that frequently co-occur with each other will form a latent semantic
type or community in \texttt{DomainNet}, with many paths of varying length between them.  
Homographs will span two or more communities.
Notice however that we do not know {\em a priori} what the communities are or even how many there are.
While there is a rich literature on community detection, many approaches require knowledge of the possible communities such as the number of communities~\cite{chakraborty2017metrics}. 
Others are parameter-free, meaning they can learn the number of communities~\cite[and others]{Henderson10}.
However, in our problem the number is not only unknown, it may be massive.
A data lake with just a modest number of tables may have many attributes representing 
a multitude of different 
semantic types (communities of values)~\cite{fortunato2010community, chakraborty2017metrics}.

What we propose in this paper is to use 
network centrality measures 
that can be defined without prior knowledge of how many communities exist, their overlap, or the distribution of attribute cardinalities.  
The intuition behind centrality measures is to capture how
well connected the neighbors of a given node are. 
We define variants of these measures appropriate for the \texttt{DomainNet} bipartite graph.
We then discuss to what extent these measures may distinguish whether a data value has a single meaning or multiple meanings (the latter being a homograph).

\smallsection{Local Clustering Coefficient as a homograph score}
The local clustering coefficient (LCC)~\cite{watts1998collective} for a given value node measures the average probability that a pair of the node's neighbors are also neighbors with each other, i.e., the fraction of value-neighbor triangles that actually exist over all possible triangles.

The LCC metric is usually defined over unipartite graphs (such as the co-occurrence graph in \Cref{fig:Example_Fig}(a)).   We use the definition of value-neighbors (recall the set of all value neighbors of a value node $u$ is $N(u)$) to generalize LCC to our bipartite graph.

The pairwise clustering coefficient of two data value nodes $v$ and $w$ is defined as the Jaccard similarity between their neighbors
$$c_{vw} = \frac{N(v) \cap N(w)}{N(v) \cup N(w)}.$$

Given a graph $G$ and a value node $u$, 
the LCC is defined as the average pairwise clustering coefficient among all the node's value neighbors:

\begin{equation}
\label{eq:local_clustering_coeff}
c_u = \frac{\sum_{v \in N(u)} c_{vu}}{|N(u)|}.
\end{equation}
The LCC of a node $u$ can be computed in time $\bigO(N(u)^2)$ 
and provides a notion of the importance of a node in connecting different communities.

\begin{hypothesis}[Homographs using LCC]
A value node corresponding to a value that is a homograph will have a lower local clustering coefficient than a value node 
with a single meaning.
\end{hypothesis}

Intuitively, we expect unambiguous values to appear with a set of values that co-occur often and thus have high LCC scores.  
This behavior should be less common for homographs, which may span values from different communities as they appear in various contexts depending on their meaning.

Despite LCC's computational simplicity, the measure as defined in \Cref{eq:local_clustering_coeff} is no more than the average Jaccard similarity between the set of attributes that a value co-occurs with.
Unfortunately, it is well-known that \emph{Jaccard similarity is biased to small sets}.
As consequence, \emph{the measure is not as effective in real data lakes} 
where attribute sizes are often considerably  skewed.
Our experiments will confirm this downside of LCC.

\smallsection{Betweenness Centrality as a homograph score}
The LCC of a node is fast to compute, but it only considers the local neighborhood of a value. 
In a data lake, the local neighborhood may not be sufficient.  
In particular, the local neighborhood may not include values that are members of the same community but happen to not co-occur.
In order to overcome these two problems (missing values in the neighborhood and attributes with very different cardinalities) we look at metrics that take a more global perspective on the network.

The \emph{betweenness centrality} (BC) of a node measures 
how often a node lies on paths between \emph{all other nodes} (not just the neighbors) in the graph~\cite{10.2307/3033543}.
One way to think of this measure is in a communication network setting where the nodes with highest betweenness are also the ones whose removal from the network will most disrupt communications between other nodes in the sense that they lie on the largest number of paths~\cite{newman2018networks}.

Consider two 
nodes $v$ and $w$.
Let $\sigma_{vw}$ be the total number of shortest paths between $v$ to $w$,
and let $\sigma_{vw}(u)$ be the number of shortest paths between $v$ to $w$ that pass through $u$ (where $u$ can be any node).\footnote{Since the bipartite graph used in \texttt{DomainNet} is not homogeneous we also examined other variations of BC such as considering only values nodes as %
end points for the examined shortest paths. 
We found that using all nodes in the BC definition provided empirically the best results for finding homographs.}
The betweenness centrality of a node $u$ is defined as follows,
where $v$ and $w$ can be any node in the graph:
    \begin{equation}\label{eq:betweenness_centrality}
        BC(u) = \sum_{v \neq u, w \neq u}\frac{\sigma_{vw}(u)}{\sigma_{vw}}.
    \end{equation}
By convention $\frac{\sigma_{vw}(u)}{\sigma_{vw}} = 0$ if $\sigma_{vw}$ (and therefore  $\sigma_{vw}(u)$) is 0.

Intuitively, a homograph appears with sets of values that do not or rarely co-occur across those sets, and thus the shortest paths between such non-co-occurring nodes would have to go through the homograph node.
Conversely, unambiguous values appear with a set of values that also co-occur a lot, and thus the shortest path between them does not unnecessarily have to go through one or a few nodes.

\begin{hypothesis}[Homographs using BC]
A value node corresponding to a homograph will have a higher betweenness centrality than a value node with a single meaning.
\end{hypothesis}

\begin{example} \label{ex:BC}
The LCC scores of the \texttt{Jaguar} and \texttt{Puma} data value nodes in \Cref{tab:jaguar_ex} are 0.36 and 0.43 respectively.
The LCC scores of the other data value nodes that appear more than once, \texttt{Toyota} and \texttt{Panda}, are somewhat higher at 0.46.
The BC scores of the \texttt{Jaguar} and \texttt{Puma} value nodes in \Cref{tab:jaguar_ex} are 0.025, 0.003 respectively.
The BC of the other value nodes that appear more than once, \texttt{Toyota} and \texttt{Panda}, are at 0.002.
Since this example only uses four small tables it does not expose the possibly different rankings between LCC and BC scores but suggests that BC, even on small graphs is more discerning.
\end{example}

\introparagraph{Complexity of BC}
Calculating the BC for all nodes in a graph is an expensive computation.
A naive implementation takes $\bigO(n^3)$ time and $\bigO(n^2)$ space ($n$ denotes the number of nodes in the graph).
The most efficient algorithm to date is Brandes' algorithm~\cite{brandes2001faster} that takes $\bigO(nm)$ time and $\bigO(n + m)$ space (for unweighted networks) where $m$ is the number of edges in the graph.
Notice that this algorithm is still expensive if the graph is dense (i.e., $m >> n$).

The high time complexity of BC motivated approximations, which usually sample a subset of
nodes from the graph and thus do not calculate all shortest paths.
One common sampling strategy is to pick nodes with a probability that is proportional to their degree (nodes with high degree are more likely to appear in shortest paths).
Riondato and Kornaropoulos~\cite{riondato2016fast} provide an approximation algorithm via sampling with offset guarantees.
Geisberger, Sanders,  and Schultes~\cite{geisberger2008better} provide an approximation algorithm without guarantees that performs very well in practice.
The complexity of the approximate BC is $\bigO(s m)$ where $s$ is the number of nodes sampled.
We chose Geisberger, Sanders,  and Schultes~\cite{geisberger2008better} to approximate betweenness centrality to benefit most from its short run-time on large graphs.

\subsection{Disambiguation Using \texttt{DomainNet}}
\label{subsec:system}

\begin{figure}[!ht]
    \centering
    \includegraphics[width=\columnwidth]{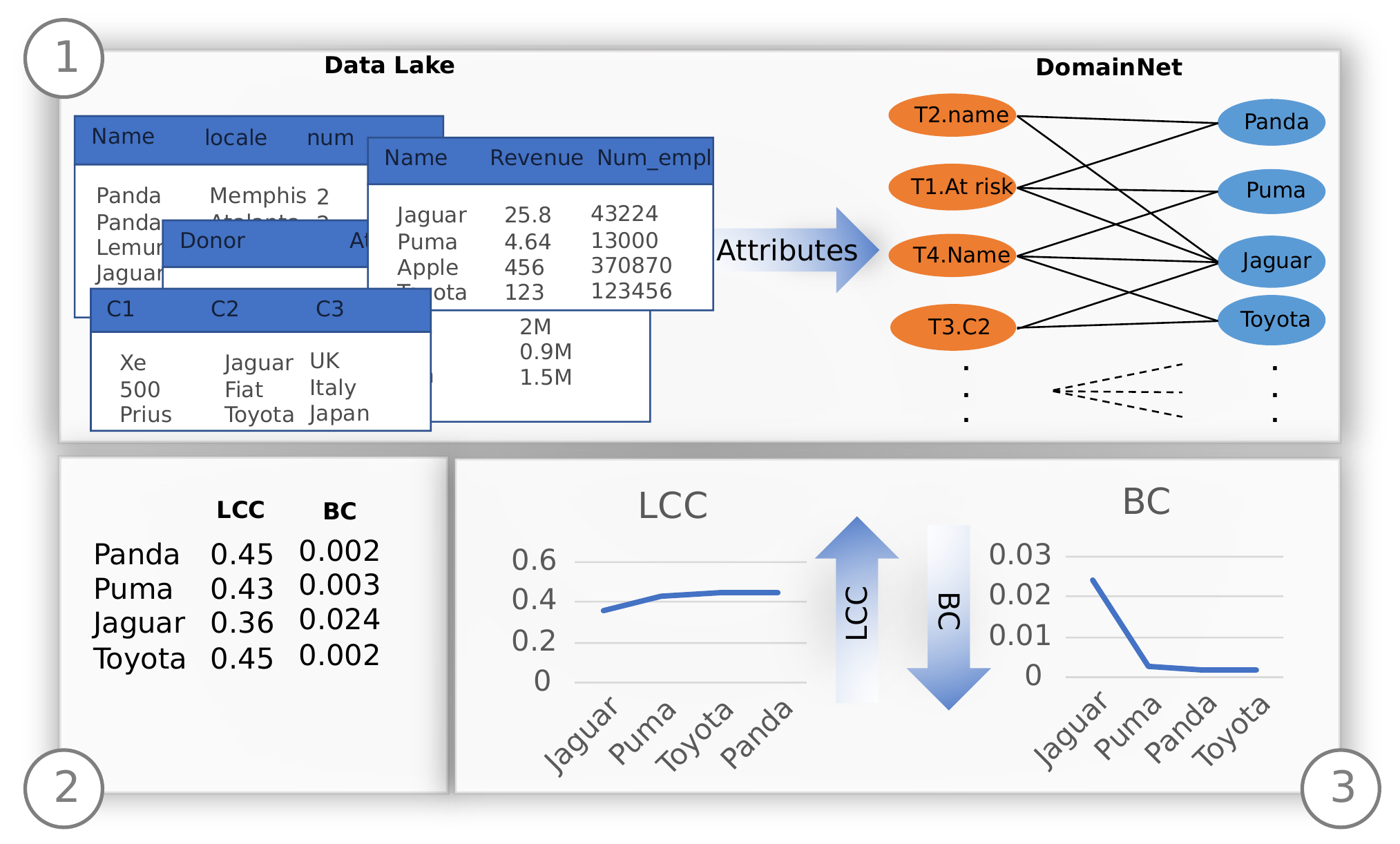}
    \caption{Disambiguation system on \texttt{DomainNet}. (1) Construct a \texttt{DomainNet} graph from a data lake. (2) Calculate BC and LCC scores for each value node in the graph. (3) Rank the scores accordingly.}
    \label{fig:workflow}
\end{figure}

In this section, we describe the implementation of an end-to-end system which allows users to disambiguate data lakes using our proposed methodology. 
Our system has three steps as illustrated in \Cref{fig:workflow}: (1) construct \texttt{DomainNet} graph; (2) calculate measures; and (3) rank measures.

{\em \texttt{DomainNet} graph construction.}
The input is a set of raw data tables from relational databases, CSV files, or any other
open data format. It is important to note that we do not require any information in regards
to types, attribute names, or the semantics of relationships between tables.
We build our bipartite graph as described in \Cref{subsec:domain_net_network}.

{\em Graph measure computation.}
Using the \texttt{DomainNet} graph constructed in the previous step, our system  computes both LCC and BC scores for each value node (\Cref{subsec:homograph_disambiguation_methodology}).
We show empirically in \Cref{subsec:SB_experiment} that BC outperforms LCC in homograph detection.  

{\em Graph measure ranking.}
Nodes are ranked by their centrality score
(ascending order for LCC measures, and descending order for BC measures) %
the top ranked data values present to a user.

\section{Dataset description}
\label{sec:setup}

Homograph detection in data lakes is a new problem and no benchmarks are available for it.
While many data lakes exist, they do not contain labels that identify the homographs.
In addition to being a hugely expensive task when done manually, homograph labeling is not a one-time effort: when the content of the data lake changes, an unambiguous value can become a homograph or vice versa.
Hence, benchmark design in this context constitutes a non-trivial contribution in itself.

We introduce the four datasets used for the evaluation of \texttt{Domain\-Net}.
The first is a new synthetic benchmark and the other three contain real data.
The second is an adaptation of the Table Union Search (TUS) Benchmark~\cite{nargesian2018table} that uses real tables from UK and Canadian open-data portals and which we adapt for our problem.
The third is a modified version of TUS, called TUS-I, where we systematically inject homographs.
The fourth, used to evaluate scalability, is a real data set from NYC Education Open Data, which was also used to evaluate a domain discovery approach~\cite{Ota+20}.

\Cref{tab:results} summarizes detailed statistics about the datasets.
For each, we list the number of tables, the total number of attributes (columns) across all tables, the number of unique values in the data lake, the total number of homographs, the range of cardinalities of any 
homograph\footnote{Recall the definition of the cardinality 
of a homograph node $v$ as $|N(v)|$, which is the number of unique data values that $v$ co-occurs with.}
(Card(H)),
and the range of the number of distinct meanings, \#M, (based on ground truth) the different homographs have across the data lake.
All datasets can be found at \href{https://github.com/northeastern-datalab/DomainNet-Datasets}{https://github.com/northeastern-datalab/DomainNet-Datasets}

\begin{table}[!ht]
\caption{
Four datasets and their statistics.
}
\label{tab:results}
\resizebox{\columnwidth}{!}{%
\begin{tabular}{lrrrrrl}
\hline
\rowcolor[HTML]{EFEFEF} 
     & \#Tables & \#Attr & \#Val & \#Hom & Card(H) & \#M   \\ \hline

SB & 13  & 39  & 17,633 & 55 & 151-1,966 &  2      \\

\rowcolor[HTML]{EFEFEF} 
TUS - I &    1,253  & 5020   &   163,860 & N/A & N/A &  N/A   \\

TUS & 1,327 & 9859 & 190,399 & 26,035 & 3-22,703 &  2-100             \\ 
\rowcolor[HTML]{EFEFEF} 
NYC-EDU &    201  &  3496  & 1,469,547 & N/A & N/A &  N/A   \\
\hline

\end{tabular}
}
\end{table}

\subsection{Synthetic Benchmark (SB)}

We designed a small fully synthetic, but real-world inspired, data lake for a systematic
validation of our approach. It consists of 13 tables generated using
Mockaroo\footnote{\url{https://www.mockaroo.com/}},
which lets the data creator specify data sources from various categories.

Each table has 1000 rows, except for two tables that contain countries and states.
We used the real numbers of countries and US states of 193 and 50, respectively. 
There are 55 data values that are homographs, e.g., Sydney (city or name), Jamaica (city or country), Lincoln (car or city), CA (country or state abbreviation), and Pumpkin (grocery product or movie title).
The benchmark along with its metadata (full list of tables and their schemas and stats) are in our github.

\subsection{Table Union Search Benchmark (TUS)}
\label{sec:TUS}

In the absence of homograph-labeled large real data lakes, we set out to find a closely related benchmark that we could adapt to our purposes.
Unfortunately, while there are many table-based benchmarks, even those for data-semantics-related problems generally proved hard to adapt.
For example, the VizNet corpus~\cite{viznet} used in semantic type detection in tables~\cite{Hul+19,Zha+19} provided ground-truth labels for only a small fraction of the columns in the repository, making ground-truth discovery of all homograph labels practically impossible.
We therefore selected the Table Union Search (TUS) benchmark~\cite{nargesian2018table}, which contains real data and provides a ground-truth mapping for \emph{each} column to the set of columns in the repository that it is unionable with.  
This enables us to automatically label all homographs.
Let $U(a)$ denote the set of columns (attributes) a given column $a$ is unionable with and notice that $a$ is always unionable with itself, hence $a \in U(a)$.
Let $A(n)$ be the set of columns (attributes) a data value $n$ appears in.
Converting the TUS benchmark into our bipartite graph representation, we can automatically label data values as ``unambiguous'' or ``homograph'' based on the unionability ground truth.
\begin{definition}[Homograph in the table union search benchmark]
\label{def:homograph_table_union_search}
	A data value $n$ is a homograph if there exist two attributes $a$ and $a'$ in $A(n)$
	such that $U(a) \neq U(a')$; otherwise $n$ is an unambiguous value.
\end{definition}

Intuitively, a data value is a homograph if it appears in at least two different columns that are not unionable (and hence have different types).
For instance, assume value \texttt{USA} appears in columns \texttt{country\_x1} and \texttt{location\_x2} in tables X1 and X2, respectively. 
If the corresponding two columns are unionable, i.e.,
$U(\textrm{country\_x1)}$ = $U(\textrm{location\_x2)}$ = $\{\texttt{country\_x1}, \texttt{location\_x2}\}$,
then we can conclude that \texttt{USA} is an unambiguous value.
In contrast, the columns containing the value \texttt{jaguar} in the zoo or donor tables   are not unionable with either the company or car model tables and hence \texttt{jaguar} would be labeled a homograph.

Based on \Cref{def:homograph_table_union_search} there are 164,364 unambiguous values
and 26,035 homographs in the TUS benchmark, suggesting homographs are very abundant in
real data lakes. 
Notice that attribute cardinalities in TUS have high skew, 
a common phenomenon in data lakes for open-data repositories~\cite{nargesian2019tutorial}.
Hence, this benchmark provides a ``stress-test'' for our approach.   How well can it deal with
both small and large cardinalities of attributes containing a homograph (in TUS these
cardinalities range from 3 to 22,703).

\subsection{TUS with Injected Homographs (TUS-I)}\label{subsec:real_data_with_injected}

Having real data is important, but we also need to understand the performance of our solution as the number of homographs in a data lake changes.
To this end, we modified the TUS benchmark as follows.
First, we removed all 26,035 homographs.
Second, we carefully introduce artificial homographs with different properties.
Since the artificial homographs are now the only ones in the data lake, we can measure how
their properties affect the detection algorithm.

A homograph is injected by selecting two different data values from two columns that are not unionable.
These original values are then replaced by a new unique value such as ``InjectedHomograph1''. 
We only replaced string values with at least 3 characters. 
In our experiments, we vary the minimum allowed cardinality of the attributes containing values replaced with an injected homograph.
We also vary the number of meanings of an injected homograph.
This allows us to evaluate the effectiveness of our approach in identifying homographs with respect to the cardinality and number of meanings of the homographs.

\section{Experimental Evaluation}
\label{sec:experiments}

The main goal of the experiments is to evaluate how well \texttt{DomainNet} performs in terms
of precision and recall for identifying the homographs in the benchmark datasets.
We are particularly interested in determining if the more expensive betweenness centrality (BC) provides significant improvement over local clustering coefficients (LCC) (\Cref{sec:approach+networks}).
Since a homograph candidate must appear in at least two different table columns, \texttt{DomainNet} pre-processes the input to remove data values that appear only once in the data lake.
As a result, the corresponding graph representation has about 3\% fewer nodes in the TUS benchmark and 30\% fewer nodes in SB.
Moreover we examine how our method scales with larger input graphs and how homographs can impact existing data integration tasks such as domain discovery.

\introparagraph{Comparison to a baseline}
There is no previous work that directly explores 
homograph detection in data lakes (\Cref{sec:related}),
and previous work on the related problem of semantic type detection and domain discovery is generally supervised, i.e., requires labeled training data.
Hence, the only suitable algorithm that we could reasonably adapt to solve our problem is the recently proposed state-of-art unsupervised domain-discovery algorithm $D^4$~\cite{Ota+20}.
We used the original code provided by the authors\footnote{The code is available at \url{https://github.com/VIDA-NYU/domain-discovery-d4}.} with its default parameter settings.
When applied to a data lake, $D^4$ assigns attributes to the discovered domains.
A natural way to identify homographs then is to identify data values that appear in more than one of those domains.  
We compare $D^4$ to \texttt{DomainNet} on the synthetic benchmark as it only contains string values.
$D^4$ discovers domains only for string data, making it ineffective on the TUS benchmark,
which contains real data with many numerical attributes.

\introparagraph{Measures of success}
We generally measure precision and recall, which are reported for the $k$ top-ranked homograph candidates identified by each of the algorithms.
By default $k$ is set to the true number of homographs in the data lake.

\introparagraph{Software implementation}
We implemented \texttt{DomainNet} in Python 3.8, using 
Networkit~\footnote{\url{https://networkit.github.io}} \cite{staudt_sazonovs_meyerhenke_2016}
to calculate exact and approximate BC scores over our bipartite graph.
This is a Python library for large-scale graph analysis whose algorithms are written in C++ and
support parallelism.
All our experiments were run on a commodity laptop with 16GB RAM
and an Intel i7-8650U CPU.

\subsection{Fully Synthetic Benchmark (SB)}\label{subsec:SB_experiment}

\begin{figure*}
    \centering
    \includegraphics[width=\textwidth]{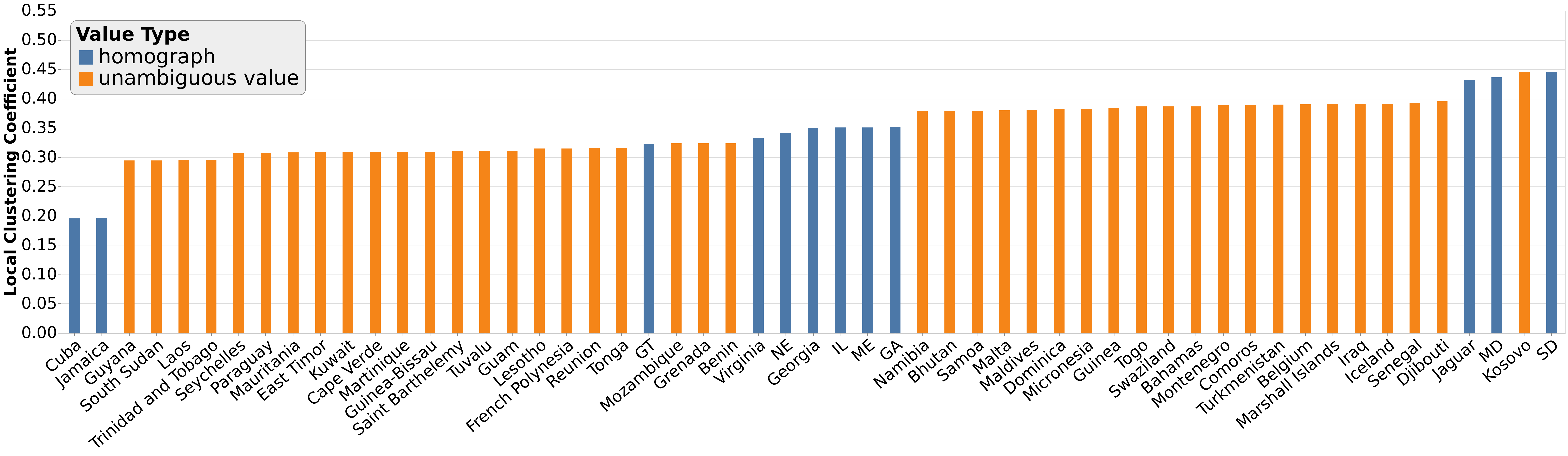}
    \caption{
    The top-55 data values with the lowest local clustering coefficients.
    Homographs are scattered throughout and do not necessarily have low LCC coefficients.
    }
    \label{fig:synthetic_LCC_topk}
\end{figure*}

\begin{figure*}
    \centering
    \includegraphics[width=\textwidth]{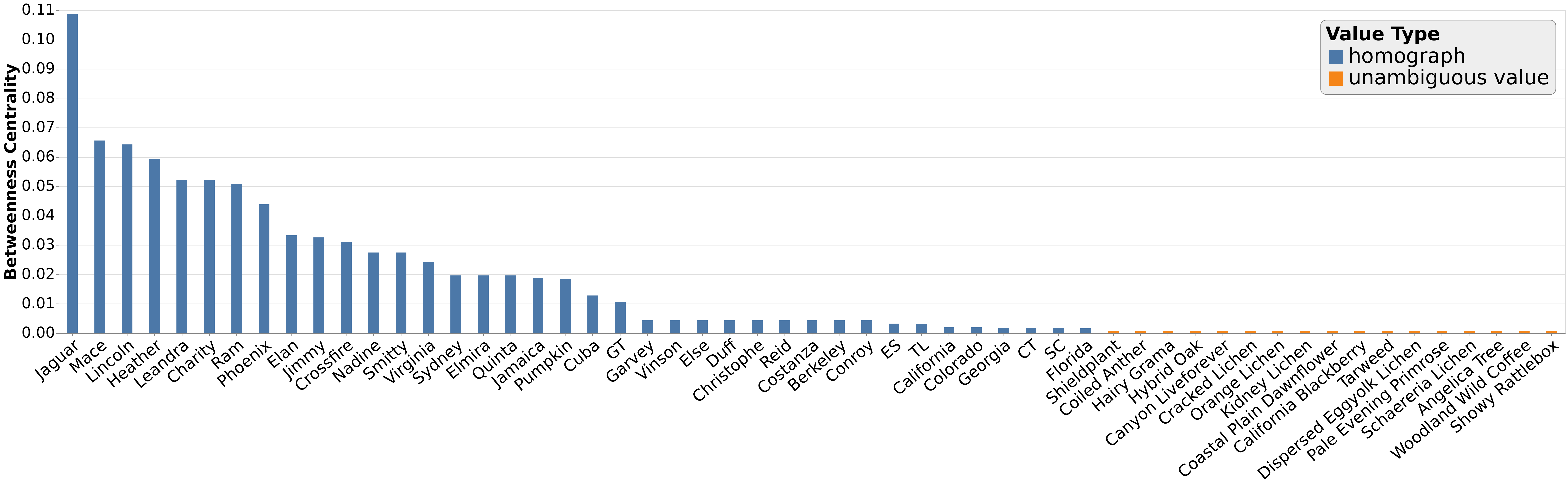}
    \caption{
    The top-55 data values with the greatest betweenness centrality scores.
    In the top-55 data values, 38 of them are homographs.
    The homographs not in the top-55 are country/state abbreviation homographs.
    }
    \label{fig:synthetic_BC_topk}
\end{figure*}

We first use the SB to compare the homograph rankings obtained using the LCC and BC measures
(\Cref{sec:approach+networks}) in order to study their ability to identify homographs.
The bipartite graph for SB is relatively small, consisting of 17,672 nodes (17,633 data-value nodes and 39 attribute nodes) and 19,473 edges.
We calculated the local clustering coefficients (LCC) and betweenness centrality (BC) for each node in the graph and examined how these scores differ between homographs and unambiguous values.

\paragraph{Which measure is better at discovering homographs?}
\Cref{fig:synthetic_LCC_topk} shows the top-55 data values based on LCC.
For LCC, lower scores should in theory indicate a greater probability of being a homograph.
Notice how more than 75\% of the top-ranked data values are not homographs, meaning that a large number of unambiguous values have smaller LCC scores than the homographs.
This is mainly caused by unambiguous values from small domains that do not co-occur often with many values in their domain.
This confirms our hypothesis from \Cref{sec:approach+networks} that LCC may not
work well when homographs appear in small domains.
In fact, the majority of the 55 homographs in the dataset have LCC scores significantly above 0.45 and so it is not necessarily true that homographs have low LCC cores.
Overall the results indicate that LCC scores do not provide an effective separation between homographs and unambiguous values.

On the other hand, the BC scores result in a vastly better top-55 result as shown in \Cref{fig:synthetic_BC_topk}.
Here 38 out of the top-55 BC scores correspond to homographs.
This is a much improved outcome over the LCC scores in \Cref{fig:synthetic_LCC_topk}.
But what happened to the remaining 17 homographs that are not in the top-55?
We noticed that the remaining 17 homographs have betweenness scores of nearly zero and they all are values corresponding to homographs that are abbreviations of country and state names.
Recall that these are the only two tables in SB with fewer than 1000 tuples, where the
state table contains only 50 tuples.
This means that the BC score for values in these small domains cannot be very large as there cannot be as many shortest paths that would pass through the homograph in question.

An explanation for the low BC scores for these homographs is the fact that there is considerable intersection between the country and state values which is not the case with other homographs (e.g., the car brands and cities intersect only on the value Lincoln and Jaguar).
This relatively large intersection also reduces the BC scores for those homographs as the number of shortest paths connecting two nodes between cities and states is much larger. 
For example, going from the country code GR to the state code MA, the shortest path could be using the homograph AL (which is for Albania/Alabama) or CA (which is for Canada/California) or any other homograph between countries and states.
As a result those homographs receive lower BC scores, because the denominator in 
\Cref{eq:betweenness_centrality} becomes large.

\paragraph{How good is previous work at finding homographs?}
As discussed earlier, we compare \texttt{DomainNet} against a competitor based on
$D^4$\cite{Ota+20}.
When applied to the SB dataset, $D^4$ discovers four domains corresponding to \texttt{Country}, \texttt{Country Code}, \texttt{Scientific Animal Name}, and \texttt{Scientific Plant Name}.
It maps the domains on 14 out of 39 table columns (attributes) in SB.
Among these 14 attributes, there are 21 of the 55 homographs.
Overall, when considering the top-55 results returned, the $D^4$-based algorithm disambiguates homographs in SB with a precision, recall, and F1-score of 38\%.
Using the BC score, \texttt{DomainNet} achieves for the top-55 results a precision, recall, and F1-score of 69\%.

\subsection{Experimental Evaluation on TUS-I}

We now study the BC-score-based version of \texttt{DomainNet} in more detail on the large real-world dataset TUS-I with the injected homographs.
Due to the cost of running BC for each node, all BC scores are approximated using 5000 samples.
\footnote{A common heuristic for the sample size is about 1-3\% of the total number of nodes in the graph. This works well in practice with sparse graphs like \texttt{DomainNet}~\cite{geisberger2008better}.
We will further test the validity of this heuristic in \Cref{subsec:scalability}.}

\begin{table}[!ht]
\centering
\caption{\% of the 50 injected homographs appearing in the top-50 results vs. cardinality of the data values replaced by the injected homograph. (Numbers are averages of 4 runs for each threshold.)}
\label{fig:injected_homograph_rank_vs_cardinality}
\resizebox{\columnwidth}{!}{%
\begin{tabular}{
>{\columncolor[HTML]{EFEFEF}}l l
>{\columncolor[HTML]{EFEFEF}}l l
>{\columncolor[HTML]{EFEFEF}}l l
>{\columncolor[HTML]{EFEFEF}}l }
\hline
Cardinality of replaced values      & $> 0$ & $\geq 100$ & $\geq 200$ & $\geq 300$ & $\geq 400$ & $\geq 500$ \\
\% of injected homographs in top 50 & 85\%     & 93.5\%     & 93.5\%     & 95\%       & 94.5\%     & 97.5\%     \\ \hline
\end{tabular}%
}
\end{table}

\paragraph{How does cardinality affect homograph discovery?}
Recall that after removing all original homographs in TUS, the TUS-I dataset only contains the homographs we methodically injected in order to study a specific effect on betweenness centrality.
We ran our experiments by randomly selecting 50 pairs of values from different domains\footnote{Different domains in the TUS benchmark context means values from columns that are not unionable with each other.} and replaced them with our 50 injected homographs.
Each experiment was repeated $4$ times with a different seed for selecting the values for replacement.
Since the number of homographs in our experiment is always $50$, in an ideal scenario the top-50 BC scores would correspond to exactly those injected homographs.

We found that cardinality has the expected impact on BC scores in terms of separating homographs and unambiguous values.
If the data values chosen for replacement have a 
not too small cardinality (i.e., they co-occur with many other values) then the BC score of their injected homograph was notably higher.
We confirmed this observation in \Cref{fig:injected_homograph_rank_vs_cardinality} where we varied the cardinality threshold for the data values chosen for replacement.
Overall, as we increased the cardinality threshold, a larger percentage of the injected homographs ranked in the top-50.
In fact, if the replaced values had a cardinality of $500$ or higher, \texttt{DomainNet} consistently ranked at least 48 of the 50 injected homographs in the top 50.  For reference, the largest attribute in TUS has 25,000 values and over half of all attributes
have more than 500 values.

\begin{table}[t]
\centering
\caption{\% of injected homographs in the top 50 according to betweenness centrality while varying the number of meanings of the injected homographs}
\label{tab:num_meanings_of_injected_homographs_vs_rankings}
\resizebox{\columnwidth}{!}{%
\begin{tabular}{
>{\columncolor[HTML]{EFEFEF}}c 
>{\columncolor[HTML]{FFFFFF}}c 
>{\columncolor[HTML]{EFEFEF}}c 
>{\columncolor[HTML]{FFFFFF}}c 
>{\columncolor[HTML]{EFEFEF}}c 
>{\columncolor[HTML]{FFFFFF}}c 
>{\columncolor[HTML]{EFEFEF}}c 
>{\columncolor[HTML]{FFFFFF}}c }
\hline
\# meanings of injected homographs & 2    & 3    & 4    & 5    & 6   & 7   & 8   \\
\% of homographs in top 50         & 97.5 & 97.5 & 98.5 & 98.5 & 100 & 100 & 100 \\ \hline
\end{tabular}%
}
\end{table}

\paragraph{How does the number of meanings of a homograph affect homograph discovery?}
In addition to varying the cardinality of the replaced values, 
we also examined how the number of meanings of the injected homographs impacts their BC-based rankings.
The number of meanings of an injected homograph is the number of values replaced for each injected homograph.
The replaced values are all chosen from different domains to ensure that the injected homographs have consistently the specified amount of meanings.
We explored injected homographs with the number of meanings in the range 2 to 8 for replaced data values with a cardinality of 500 or higher.
\Cref{tab:num_meanings_of_injected_homographs_vs_rankings} shows that as we increase the number of meanings, \texttt{DomainNet} becomes better at discovering them.
This is consistent with our intuition for betweenness centrality since homographs with more meanings are more likely to be hub nodes that connect multiple sets of nodes with each other in our bipartite graph representation of the data lake.

\subsection{Homographs in TUS Benchmark}

\begin{figure}[!ht]
    \centering
    \includegraphics[width=\columnwidth]{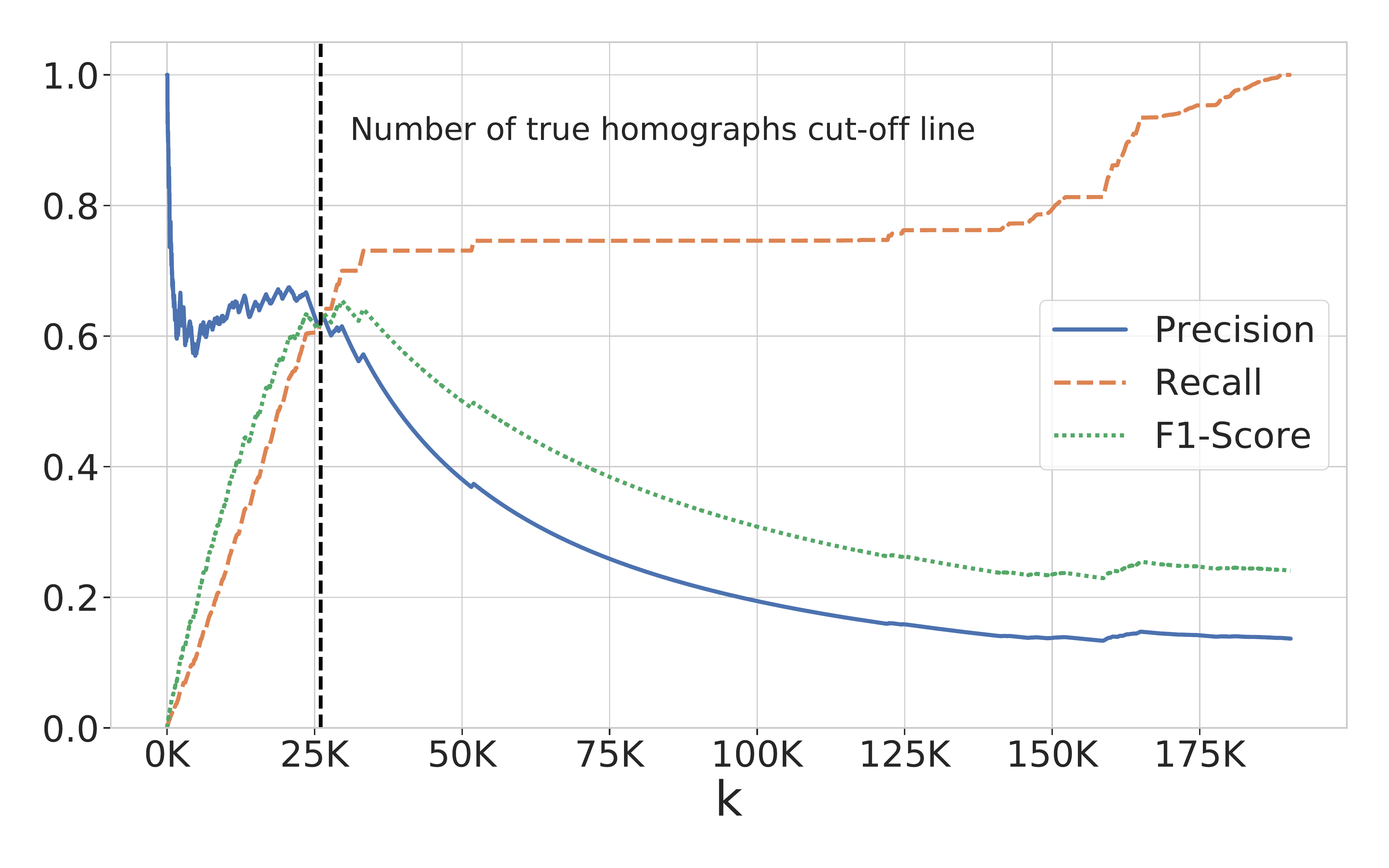}
    \caption{
    Top-k evaluation on the TUS dataset. The vertical line at $k$=26,035 denotes the number of true homographs in the dataset.
    }
    \label{fig:TUS_evaluation}
\end{figure}

Lastly, we explore the performance of \texttt{DomainNet} with betweenness centrality on the real TUS dataset with its 26,035 real homographs.
Since the number of homographs is large, we not only report precision, recall, and F1-score for the top-26,035 results, but for all top-$k$ with $k$ from 1 all the way to the number of nodes in our graph, i.e., 190,399.
We do not compare against the $D^4$-based algorithm for homographs, because $D^4$ operates only on string attributes, and given the large number of numerical attributes the $D^4$ coverage will be even lower than in SB (where it only finds domains for 14 out of 39 attributes).

\paragraph{How does our approach perform on a real open-data benchmark?}
\Cref{fig:TUS_evaluation} shows the summary of our top-$k$ evaluation results.
Notice that for relatively small values of $k$ such as $k=200$ our method can identify homograph values with high precision (0.89).
Naturally, as we increase $k$ precision decreases and recall increases.
At $k=26,035$ (vertical line in \Cref{fig:TUS_evaluation}), which is the number of true homographs in the TUS benchmark, we achieve a precision, recall and F1-score of 0.622.
The highest F1-score occurs at $k=29,633$ where precision, recall, and F1-score are 0.615, 0.7 and 0.655, respectively.

It is important to emphasize that our approach is completely unsupervised and does not
assume any external knowledge about the tables or their values.
Existing state-of-the-art methods that tackle data integration tasks as described in \Cref{sec:related} cannot be readily used for homograph identification or their coverage is severely limited (e.g., knowledge-based approaches like AIDA \cite{yosef2011aida}). 

Below we report the top-10 values and their BC scores from the TUS benchmark.
\begin{itemize}
    \item  ``Music Faculty'' $\rightarrow$      0.00064  
    \item    ``Manitoba Hydro''  $\rightarrow$ 0.00045  
    \item    ``50'' $\rightarrow$      0.00029 
    \item    ``1800ZZMALDY2''  $\rightarrow$    0.00028   
    \item    ``.'' $\rightarrow$      0.00027
    \item    ``Conseil de développement'' $\rightarrow$  0.00025
    \item    ``125'' $\rightarrow$  0.00023 
    \item    ``2'' $\rightarrow$  0.00022
    \item    ``Biomedical Engineering'' $\rightarrow$  0.00022 
    \item    ``SQA'' $\rightarrow$  0.00016
\end{itemize}
All 10 data values are homographs based on the ground truth.
Notice that from a natural-language perspective these 10 values do not seem to be homographs, but a closer look at the data revealed good reasons why they were labeled as homographs.
For example the value \emph{Music Faculty} appears in two distinct contexts: as a geographic location/landmark in transportation-related tables as well as a department in university-related tables.

The value with the fifth-highest BC score is the period character.
This may seem bizarre, but the period is used extensively as a null replacement in a large variety of tables and thus it acts as a homograph with a very large number of meanings.
Finally, notice that we identify numerical values such as 50, 125 and 2, which appear in a variety of contexts such as addresses, identification numbers, quantity of products, etc.
Numerical values are traditionally difficult to deal with in many data-integration tasks,
hence being able to identify some of them in a completely unsupervised manner is a notable step toward better coverage for numerical values.

\subsection{Scalability}
\label{subsec:scalability}

As discussed in \Cref{subsec:system}, 
Step 1 (graph construction) and Step 2 (centrality measure computation) are the most computationally expensive in our approach.
In this section, we examine empirically the scalability of these steps.

The time to construct our bipartite graph is dependent on how long it takes to scan all
input tables, which is a relatively fast operation.
For example, the bipartite graph for the TUS dataset takes about 1.5 minutes to construct, which is how long it takes to read through each table in the dataset.

The runtime of Step 2 depends on the graph measure used.
LCC is a local measure that is efficient to compute, but as we demonstrated in \cref{subsec:SB_experiment} it is not as effective in finding homographs as BC is.
Computing the LCC score for every node in the TUS dataset takes 4 seconds.
For the global measure BC, since we are more interested in the score rankings rather than the scores themselves, approximating BC via sampling can significantly decrease the runtime without compromising quality.

\begin{figure}[!ht]
    \centering
    \includegraphics[width=\columnwidth]{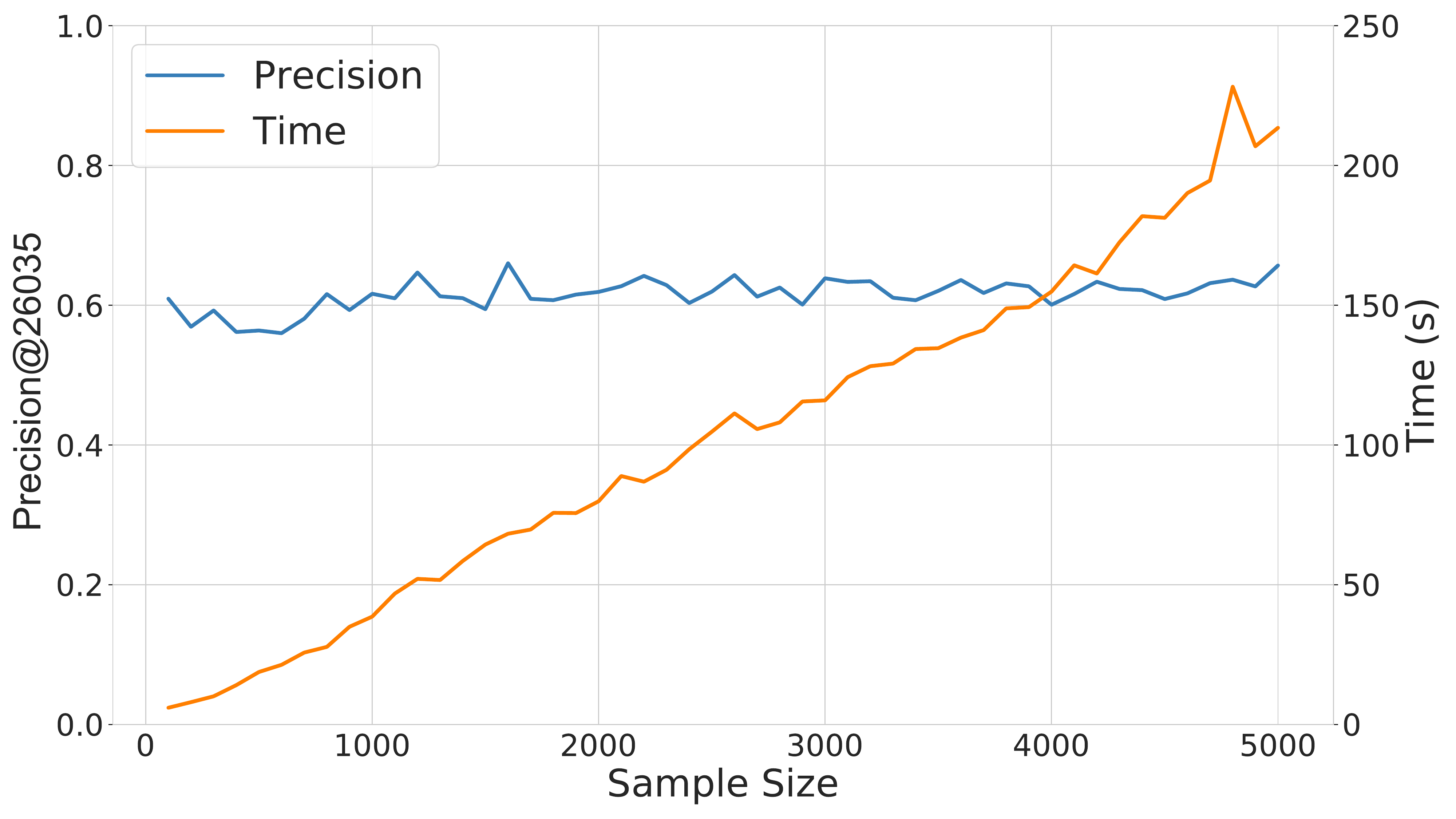}
    \caption{
        Precision at $k$ (where $k$ is the number of homographs in the dataset) and execution time at various sample sizes for approximate BC on the SB and TUS datasets.
        Exact BC on TUS took 150 minutes with a precision of 0.631.
    }
    \label{fig:scalability}
\end{figure}

In \Cref{fig:scalability}, we examine how precision and runtime vary as we change the number of samples used for the approximate BC algorithm~\cite{geisberger2008better} on the TUS benchmark.
Even for a small sample size (e.g., 1000), precision stabilises at 0.6.
Notice that 1000 samples correspond to around .5\% of the nodes in the TUS graph and it takes about 40 seconds for the algorithm to complete.
The BC approximation has a complexity of $\bigO(sm)$ where $s$ is the number of nodes sampled and $m$ the number of edges in the graph.
Based on the literature and testing on our graphs we found that sampling $1\%$ of the nodes provides a good approximation of BC that is very consistent with the score rankings produced by the exact BC computation.

We also considered a bigger data lake to further test execution times---the
NYC education open data dataset as used in $D^4$ \cite{Ota+20}.
The bipartite graph representation of that dataset has roughly 1.5M nodes and 2.3M edges which is an order of magnitude larger than the bipartite graph for the TUS dataset.
The graph was constructed in 3.5 minutes and the BC scores for every node were computed in 27 minutes using approximate BC on 1\% of the nodes ($\sim$15K nodes).

\begin{figure}[!ht]
    \centering
    \includegraphics[width=\columnwidth]{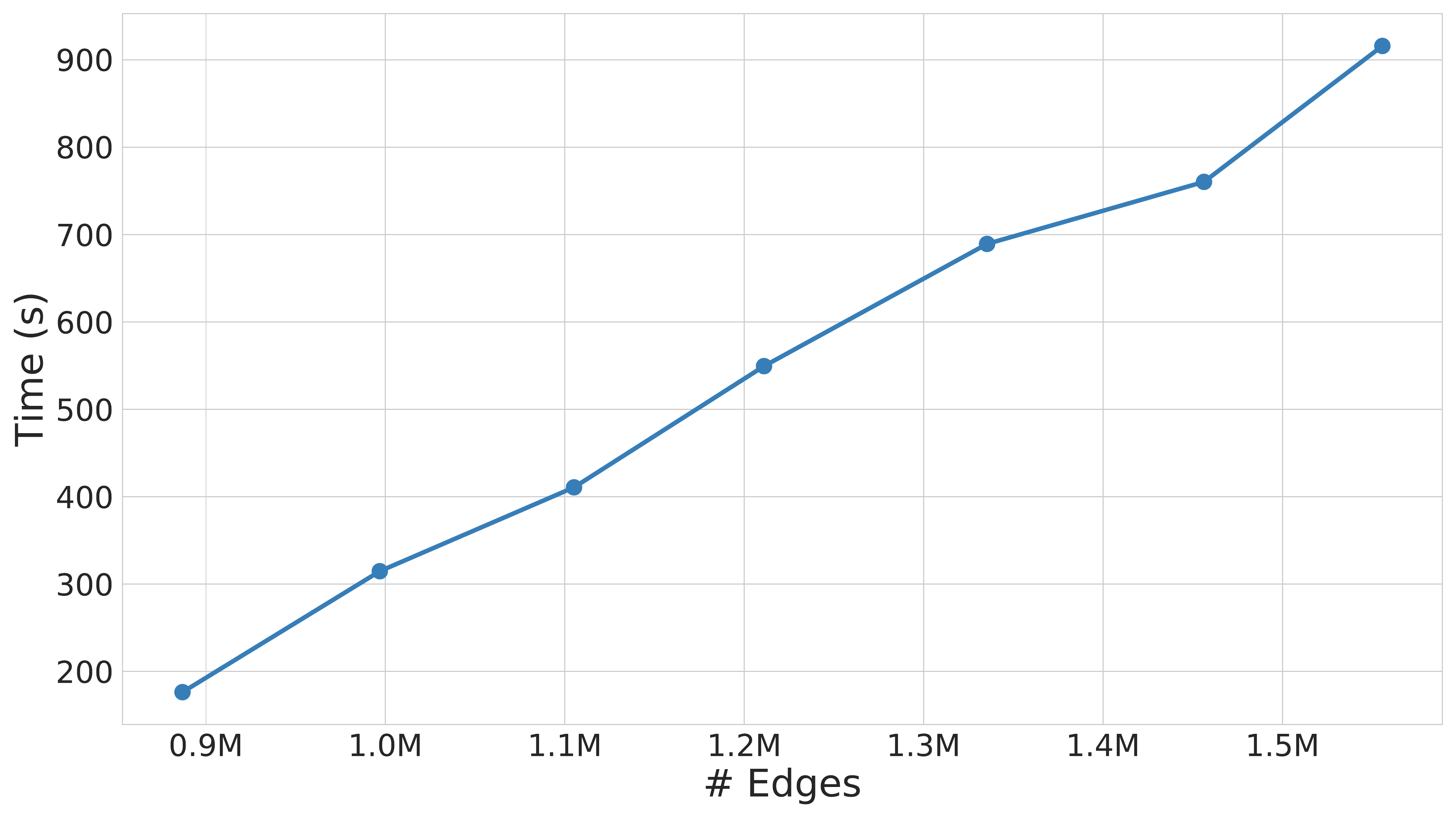}
    \caption{
        Runtime of approximate BC for various sized subgraphs based on the NYC education dataset.
    }
    \label{fig:scalability_subgraphs}
\end{figure}

To examine how runtime scales with graph size we extracted random subgraphs\footnote{The  subgraphs were constructed by randomly selecting an attribute node and adding all its connecting value nodes. We repeat by selecting another attribute node until the subgraph reaches the desired size (within some margin)} of various sizes from the bipartite graph used for the NYC education dataset.
We ran approximate BC for each graph by sampling $1\%$ of its nodes and measured the runtime.
\Cref{fig:scalability_subgraphs} shows that runtime increases linearly with graph size (i.e., number of edges)  which is in accordance with the $\bigO(sm)$ complexity of the approximate BC algorithm.
.

\subsection{Impact of homograph discovery on $D^4$}\label{subsec:impact_of_homographs}

As shown in \Cref{tab:results} the number of homographs in a real data lake can be large.
To further understand the impact of homographs on existing approaches, we consider
the task of domain discovery and examine how knowing homographs {\em a priori} can benefit them.

\begin{figure}
    \centering
    \includegraphics[width=\columnwidth]{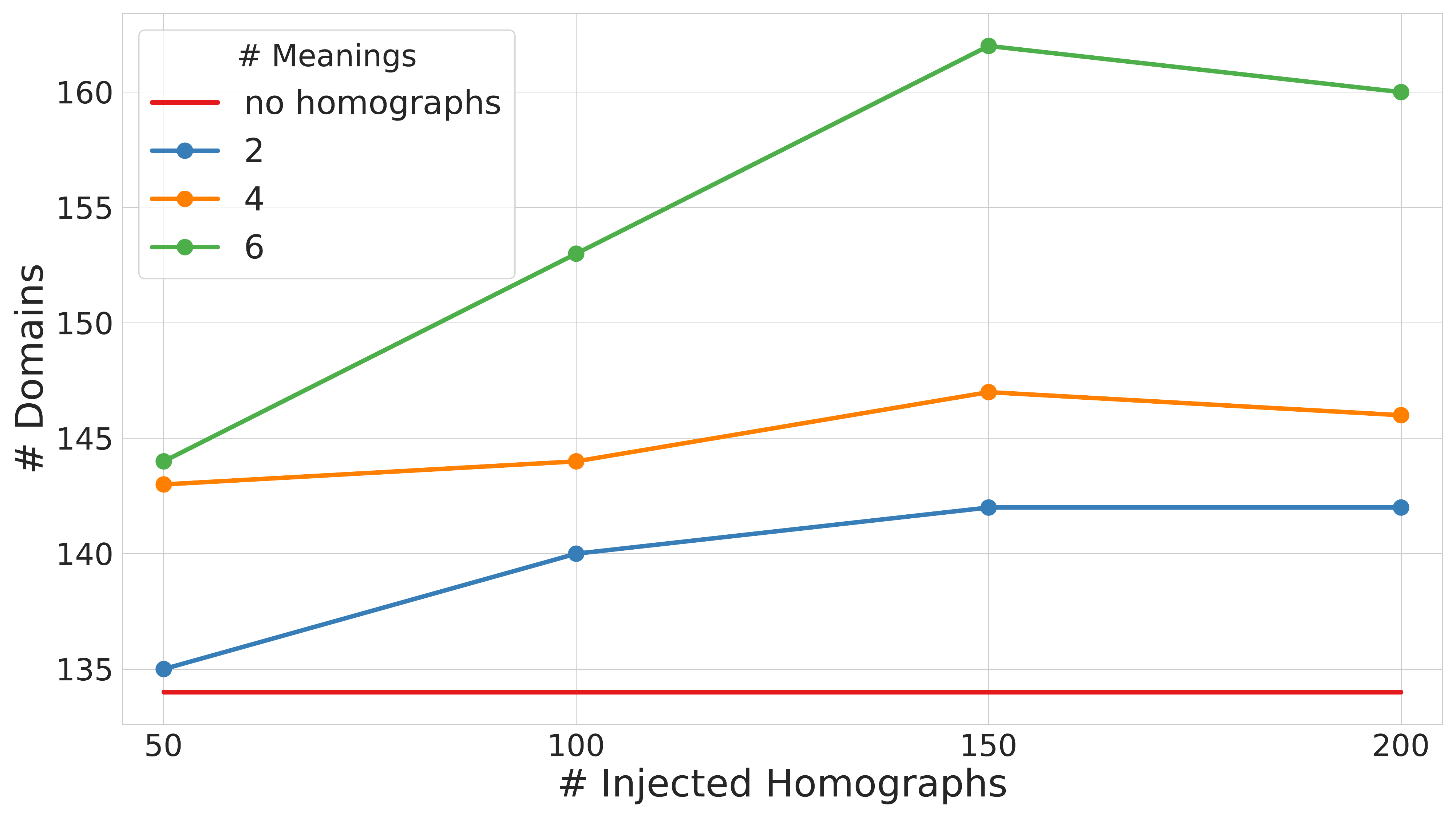}
    \caption{
        Number of domains found by $D^4$ over different TUS-injected datasets.
        The horizontal red line shows the number of domains found when no homographs were present in the dataset.
    }
    \label{fig:TUSonD4}
\end{figure}

We report the results of five different runs of $D^4$ in \Cref{fig:TUSonD4}.
The plots show the number of domains found by $D^4$ (y-axis) as we vary the number and meanings of the injected homographs. 
To be fair in the comparison and to understand the impact of homographs on the domain discovery task, we use the TUS-I benchmark.
We first ran $D^4$ over the dataset without homographs and then over the same dataset
with injected homographs.
More specifically, we injected 50, 100, 150 and 200 homographs with 2, 4 and 6 meanings.
In all the above configurations the dataset always had 68 domains based on the ground truth.
The horizontal line in \Cref{fig:TUSonD4} shows that $D^4$ returns 134 domains for TUS-I with no homographs.
The difference in the number of domains based on the ground truth and $D^4$'s results is due to the nature of the TUS benchmark~\cite{nargesian2018table} as it is created from a set of large real open data tables that were randomly sliced vertically and horizontally.
Consequently, in some cases the columns originating from the same table no longer share any values, causing $D^4$ to discover more domains than there are based on ground truth.

As we increase the number and meanings of the injected homographs, $D^4$ returns even more domains leading to lower accuracy.
$D^4$'s output provides statistics about the maximum and the average number of domains assigned to a column. 
In the TUS-I with no homographs, that maximum is 2 and the average is almost 1 (i.e., 1.031) and it increases with the number of homographs.
With 200 homographs the maximum is 4 and the average is 1.04. 
We also ran $D^4$ on the TUS-I with 5000 injected homographs, to simulate a dataset with a large proportion of homographs as in the TUS benchmark.
The maximum domains per column is 22 and the average is 1.7 with a total of 371 domains found.
The presence of homographs is negatively affecting $D^4$ and causing it to erroneously assign larger numbers of heterogeneous domains to attributes as the number of homographs increases.
Homograph discovery therefore is an important step that can be executed before domain
discovery to improve its performance.

\section{Conclusion and Future work}
\label{sec:conclusion}

We presented \texttt{DomainNet}, a method for finding homographs in data lakes.
To the best of our knowledge, this is the first solution for disambiguating data values in data lakes.
Notably, our approach does not require complete or consistent attribute names.

We showed that a measure of centrality can effectively separate homographs from unambiguous values in a data lake by representing tables as a network of connections between values and attributes.

We compared against an alternative approach using $D^4$ to identify the semantic domain (type) of attributes~\cite{Ota+20} and labeling a value a homograph if it appears in more than one domain.  
Our direct computation of homographs has significantly better precision and recall than the domain-discovery approach.
This seems to be due to $D^4$ at times placing homographs into a domain represented by their most popular meaning and the fact that $D^4$ does not find domains for every attribute.
When we inject homographs into real data, \texttt{DomainNet} is robust to the number of meanings of the homographs, reliably finding homographs with even better accuracy as the number of meanings increases.
We also demonstrated the importance of homograph detection by showing that the presence of homographs can have considerable impact on 
existing semantic integration tasks (specifically, domain discovery).

In a benchmark created from real data, our method provides a clear separation with high precision of homographs from values that are repeated, but always with the same meaning.
The accuracy is influenced by the cardinality of the homograph (i.e., the number of data values with which the homograph co-occurs).
When this number is too small, the
bipartite graph representation 
is not always sufficient to effectively identify all homographs.
In our experiments, the accuracy dropped from 97\% to 85\% as we reduced the cardinality of homographs.  

The homographs we discover on real data include phrases with multiple meanings (e.g., \texttt{Music Faculty} referring both to a geographic location and to a University unit).
They also include null values (e.g., a dot ``\texttt{.}'' can indicate unknown/missing $X$ where $X$ varies in different contexts) and data errors (e.g., \texttt{Manitoba Hydro}, an electric company, is placed in the wrong column \texttt{Street Name}).
In NLP, previous work on disambiguation primarily focuses on the disambiguation of words and named-entities.
Our method is purely based on co-occurrence information
and does not discriminate between different types of homographs. 
In fact, we provide the first approach to disambiguate numerical values in tables (e.g. \texttt{25} can be a street number or an ID number).

Identifying homographs from tables in a completely unsupervised manner can play an important role in improving other data-lake analysis tasks.
Specifically, we are considering how to determine if a homograph is an error, e.g., the value has been placed in the wrong cell.
With such knowledge, we can help not only identify such errors, but clean them as well.
We also believe that our homograph metrics can improve supervised semantic type detection such as Sherlock~\cite{Hul+19} or SATO~\cite{Zha+19}.

In this context, it will also be important to determine the number of distinct meanings
of a homograph.
Our approach is motivated by work on community detection where a community represents a meaning for a value (e.g., animal or car model).
Hence we are investigating the role of community detection algorithms on discovery
of meanings of values in data lake tables.
Notice that in this problem, we do not know {\em a priori} what the communities are or even how many there are.
Non-parameterized community detection algorithms can be used to discern the number of meanings of homographs.
However, 
innovation is needed for homographs with large numbers of meanings (such as null equivalents)~\cite[and others]{Henderson10}.

To the best of our knowledge there are no available benchmarks for homograph detection. Our synthetic benchmark (SB) and our benchmarks TUS and TUS-I (that use real open data tables~\cite{nargesian2018table}) are the first open benchmarks in this area.

In order to design a robust and completely unsupervised solution that scales to
large data lakes, we have quite deliberately limited \texttt{DomainNet} to
use only value co-occurrence information in table columns,
ignoring additional structural information like co-occurrence of values in the same row.
Our goal was to explore how much this information alone reveals about data value semantics.
Given our strong positive results, we believe our metrics should become an important feature that could be used in other problems that involve understanding or integrating tables.
An important open problem is to extend \texttt{DomainNet} to {\em collectively} resolve ambiguous metadata and data, perhaps using probabilistic graphical models that have been applied to collectively resolving multiple types of entities at once~\cite{Kou+19} and to collectively resolving data and metadata inconsistency in schema mapping~\cite{DBLP:journals/tkde/KimmigMMG19}.

\smallsection{Acknowledgments}
This work was supported in part by
the National Science Foundation (NSF) under award numbers 
IIS-1956096 and
CAREER IIS-1762268.

\bibliographystyle{abbrv}
\bibliography{main}

\newpage
\appendix

\end{document}